\documentclass[12pt,a4paper]{article}
\pdfoutput=1
\usepackage{jheppub}
\usepackage[utf8]{inputenc}
\usepackage{xcolor}
\usepackage{epstopdf}
\usepackage{graphicx}
\usepackage{epsfig}
\usepackage{dcolumn}  
\usepackage{bm}    
 \usepackage{caption}
 \usepackage{subcaption}
\usepackage{amssymb} 
\usepackage{amsmath,bm}
\usepackage{amsfonts}  
\usepackage{amsmath}  
\usepackage{slashed}  
\usepackage{enumitem}
\usepackage[mathscr]{euscript}
\usepackage{tabu}
\usepackage{epsfig}
\makeatletter
\g@addto@macro\bfseries{\boldmath}
\makeatother

\def\l1{{{1-loop}}}

\def\n1{\Bigg|_{n=1}}

\def\n{{(n)}}

\usepackage{bookmark}
  \title{\textbf{\textsf{Hyperbolic cylinders
  and entanglement entropy: gravitons, higher spins, $p$-forms}}}
  \author{Justin R. David, Jyotirmoy Mukherjee}
\affiliation{\vspace{.1cm} Centre for High Energy Physics, \\ Indian Institute of Science,\\
C. V. Raman Avenue, Bangalore 560012, India.}
\emailAdd{justin@iisc.ac.in, jyotirmoym@iisc.ac.in}
\abstract{ 
We show that the  entanglement entropy of $D=4$ linearized gravitons across a sphere 
recently computed   by Benedetti and Casini coincides
with that obtained  using the  Kaluza-Klein tower of traceless transverse 
massive spin-2 fields
 on $S^1\times AdS_3$. The mass of the constant mode on $S^1$ saturates 
the Brietenholer-Freedman bound in $AdS_3$. 
This condition also ensures that the  entanglement entropy  of higher spins
determined from partition functions
on the hyperbolic cylinder
coincides with their recent conjecture. 
Starting from the action of the 2-form on $S^1\times AdS_5$ and fixing gauge, we evaluate 
the entanglement entropy  across a sphere  as well as the dimensions of the corresponding
twist operator. 
We demonstrate that
the conformal dimensions of the 
corresponding twist operator  agrees with that obtained using the expectation value of 
the stress tensor 
on the  replica cone.  
For conformal $p$-forms in even dimensions  it  obeys the 
expected relations with the coefficients determining the  $3$-point function of the 
stress tensor of these fields.  
}
 
\begin{document}

\maketitle

\section{Introduction}
\label{sec:intro}

Entanglement entropy of a sub-region in quantum field theories 
is an important quantity. 
For conformal field theories in $2$ dimensions, the coefficient of logarithmic term 
is proportional to the central charge or the degrees of freedom of the system. 
A similar statement holds true for conformal field theories in higher dimensions. 
Entanglement entropy has found applications in topics that encompasses  different
areas of theoretical physics from black hole physics and holography to phase transitions in 
condensed matter theory. 
In spite of this, it is a difficult quantity to evaluate even 
for free theories. If the theory admits a gauge symmetry then one needs to define the algebra
of gauge invariant observables in a sub-region  to evaluate the entanglement entropy.
Entanglement entropy in a theory of gravity or  a theory with higher spin fields also suffers from this 
problem due to the difficulty of defining what is a  subregion. 

Recently  Benedetti and Casini  \cite{Benedetti:2019uej} 
evaluated the entanglement entropy across the $2$-sphere
of free linearized gravitons in $D=4$ dimensions. 
This theory can be considered as a quantum field theory of spin-2 particles and it can be treated just 
as the Maxwell field. After fixing gauge and by canonically quantising the theory in spherical co-ordinates
they obtain the following result for the logarithmic coefficient
\begin{equation}\label{casbenresul1}
S_{\rm EE} = \frac{61}{45} \log (\epsilon), 
\end{equation}
where $\epsilon$ is the dimensionless cutoff. Furthermore using the observations from the spin-2
case,  they conjectured that the logarithmic coefficient for the entanglement entropy of free higher 
spin-$s$ particles across a sphere  is given by 
\begin{equation}\label{casbenresul2}
S_{\rm EE} = \frac{1+ 15 s^2}{45} \log(\epsilon). 
\end{equation}

Setting $s=1$  in (\ref{casbenresul2}) reproduces the logarithmic coefficient of the free Maxwell field. 
However this contribution does not capture the edge modes 
\cite{Donnelly:2014fua,Huang:2014pfa,Donnelly:2015hxa}
or the contribution of the electric centre
\cite{Soni:2016ogt,Moitra:2018lxn}. 
Therefore,  the entanglement entropy given in (\ref{casbenresul1}) of the gravitons
also does not include these subtle modes.  For the purposes of the present paper we 
ignore such contributions. 
The Maxwell theory is conformal and indeed, its logarithmic 
coefficient   can be obtained from studying thermodynamics in 
deSitter space which is conformally related to causal development of the 2-sphere.
Surprisingly  it was noted in \cite{Dowker:2019zva}, 
 that the logarithmic  coefficients in (\ref{casbenresul2}) can also be obtained from 
the thermodynamics of spin-$s$ particles in deSitter space using results of earlier work
\cite{Dowker:1983nt}. 
This is surprising because the spin-2 or the higher spin theory is not conformal and a local 
stress tensor for these theories is not well defined.

In this paper we wish to understand this  logarithmic coefficient for  gravitons and 
the conjecture for the higher spin particles further.  It would be satisfying if a path integral 
approach to evaluating entanglement entropy can be extended to these fields. 
For a conformal field theory in $D$ dimensions, the  
R\'{e}nyi entropy/entanglement entropy can be obtained by evaluating 
the partition of the theory on  the hyperbolic cylinder, $S^1\times AdS_{d}$ with $d+1 =D$
\cite{Casini:2011kv}. 
Therefore given the action of the theory on the hyperbolic cylinder one could perform the path 
integral and evaluate the entanglement entropy from the partition function. 
Though, the spin-2 theory is not conformal invariant, we wish to purse this 
this approach  to study the logarithmic coefficient for the spin-2  field. 

Before we proceed, as a warm up exercise we consider 
the Maxwell action on $S^1\times AdS_3$.  
We adapt the methods of gauge fixing developed in \cite{Fradkin:1983mq,David:2009xg,Gaberdiel:2010ar} for this theory and show that the 
partition function of the Maxwell theory on 
$S^1\times AdS_3$ reduces to a 
 Kaluza-Klein  tower  of the  transverse vector  on $AdS_3$. The mass of the 
 constant mode on $S^1$  saturates the Brietenholer-Freedman bound of the spin-1 field on 
 $AdS_3$, an observation which was missed earlier and which plays and important role
 in the higher spin generalisation. 
 Evaluating the entanglement entropy, we find that it precisely 
 coincides with the $s=1$ result of (\ref{casbenresul2}). 
 We also evaluate the weight of the co-dimension 2
  twist operator 
 localized on the entangling surface from the partition function. This agrees with that obtained
 using the expectation value of the stress tensor on the replica cone.

 To consolidate these observations, we also study the conformal  $2$-form in $6$ dimensions. 
 We start with the action of the $2$ form on $S^1\times AdS_5$, fix gauge and show that
 the partition function reduces to that a Kaluza-Klein of tower of co-exact $2$-forms on 
 $AdS_5$. Indeed again, the mass of the constant mode on $S^1$ saturates the 
 Brietenholer-Freedman bound. 
 For the 2-form we also evaluate the conformal dimensions of the co-dimension two 
 twist operator
 by evaluating the expectation value of the stress tensor on the replica cone generalising 
 the methods of \cite{Candelas:1977zza}. 
 The result agrees precisely with that obtained from the partition function.

 The  detailed study  of  the $U(1)$ and  the 2-form theory
  shows that the entanglement entropy and the 
 weight of the co-dimension twist operator 
 are integral transforms of the Plancherel measure of the corresponding field on 
 $AdS_5$.  We use this observation to obtain the  putative entanglement of
conformal $p$-forms in even dimensions to $D=14$. 
 At every instance we perform  consistency checks on the results. One non-trivial check
 is that the dimensions of the twist operator obey the expected relations \cite{Hung:2014npa} with the 
 parameters determining the three point function of the stress tensor 
 of the theory of conformal $p$-form.

 The theory of linearized gravitons in flat space is not conformal invariant, furthermore 
 the hyperbolic cylinder $S^1\times AdS_3$  is not a solution to the Einstein equations with 
or without a cosmological constant. These reasons prevent us from directly considering 
the theory of linearised gravitons on $S^1\times AdS_3$. 
We instead consider the gauge invariant  partition function of linearised gravitions  on 
$S^1\times R^3$ and introduce a principle to 
turn on curvature in the spatial directions.  This leads us the theory of a Kaluza-Klein tower
of traceless transverse spin-2 fields on $S^1\times AdS_3$. 
The mass of the constant mode on $S^1$ saturates the  Brietenholer-Freedman bound. 
Evaluating the entanglement entropy from this partition function we see that the logarithmic 
coefficient precisely coincides with that in (\ref{casbenresul1}). 
We demonstrate that this partition function is scale invariant and depends only 
on the ratio of the radius of $S^1$ and $AdS_3$. 
We  repeat  the same analysis on the spin-$s$ field and 
 obtain (\ref{casbenresul2}) for the entanglement entropy. 
 The  sphere is conformally related to the space $S^1\times AdS_3$
 As a consistency check on our results we show that the difference of the entanglement 
 entropy evaluated from the partition function of the Kaluza-Klein tower 
 transverse spin-$s$ fields at the Brietenholer-Freedman bound and that evaluated from 
 $S^1\times AdS_3$ and that from spin-$s$ fields precisely agrees with the edge mode partition 
 function of spin-$s$ fields evaluated recently by \cite{Anninos:2020hfj}. 
 This check also shows that the just as in the spin-1 case, the hyperbolic cylinder method 
 misses the edge mode contributions to entanglement entropy. 
 Finally we provide an argument from the form of the Plancherel measure 
 of spin-$s$ fields on $AdS_3$ why the entanglement entropy from  the partition function 
 on $S^1\times AdS_3$ must agree with that conjectured by \cite{Casini:2011kv}.

 The organisation of this paper is as follows. 
 In section  \ref{spherhyper}, 
 after a brief review 
 we revist the Maxwell theory on 
  on $S^1\times AdS_3$ and  evaluate the R'{e}nyi entropies and 
 twist operator dimensions.  
  In section \ref{antiten}  we consider the action of the  2-form in $D=6$, and evaluate the gauge 
 invariant partition function  on $S^1\times AdS_5$. This is then 
 used to evaluate the R\'{e}nyi entropies  and the twist operator dimensions. 
 We show that the twist operator dimensions coincides with that obtained in  
 from the expectation values of the stress tensor of the 2-form on the cone. 
 We use these observations and extend the analysis to conformal forms 
 in even $D\leq 14$ dimensions. 
 In   section  \ref{tenfield} we show that the entanglement entropy obtained from the 
 partition function of a Kaluza-Klein tower of gravitons on $S^1\times R^3$ at the 
 Brietenholer-Freedman  bound. 
 coincides precisely with \ref{casbenresul1}. 
 We then extend this to higher spin field in $D=4$ . As a consistency check we show 
 that the edge modes of the higher 
 spin fields using our approach precisely coincides with that obtained in  \cite{Anninos:2020hfj}.
 For completeness evaluate the entanglement entropy from 
 higher spin partition functions to all 
  and all even $D\leq 14$. 
 Section \ref{conclude}  contains our conclusions. 
 The appendices contain various details which are necessary  for completeness and 
to arrive at the conclusions in the main text.   

\section{Spherical entangling surface and hyperbolic cylinders} \label{spherhyper}

Consider  a conformal  field theory in $D$-dimensions and let $\Sigma$  be the 
$D-2$-dimensional sphere of radius $R$ at a given time. 
Let the causal development of this region be denoted by ${\cal D}$. 
We briefly recall the argument put forward in \cite{Casini:2011kv}  to evaluate the 
entanglement entropy obtained by  tracing over the region  outside the $D-2$-dimensional sphere. 
It was shown in \cite{Casini:2011kv} that the entanglement entropy of the sphere   is equal to the 
 Von-Neumann entropy  of the density matrix  of the vacuum on 
${\cal D}$ which we denote by $\rho_{\cal D}$.  
Furthermore  a conformal transformation relates the density matrix  of the theory on ${\cal D}$ 
to the thermal density matrix on the  hyperbolic cylinder
 $\mathbb{R}\times AdS_{d}$ where $d+1 = D$,  with temperature
$T = \frac{1}{2\pi R}$. 

To be explicit, the Minkowski space is given by the metric
\begin{equation} \label{mink}
ds^2 = - dt^2 + dr^2 + r^2 d\Omega_{D-2}^2. 
\end{equation}
Here $d\Omega_{d-2}^2$ is the metric on the $D-2$ unit sphere.  The entangling sphere  $\Sigma$ is
at radius $R$. 
The causal development of this sphere  can be related to the hyperbolic cylinder using the 
following co-ordinate 
transformation 
\begin{eqnarray}
t = R \frac{\sinh  \frac{ \tilde \tau }{R}  }{ \cosh u + \cosh \frac{\tilde \tau}{R}  }, \qquad\qquad 
r = R \frac{\sinh u }{ \cosh u + \cosh \frac{\tilde \tau}{R} }.
\end{eqnarray}
The Minkowski metric  in  (\ref{mink})  then transforms to 
\begin{eqnarray}
ds^2 &=& \hat \Omega^2
 [  - \frac{d\tilde \tau^2}{R^2}  + ( du^2 + \sinh^2 u d\Omega^2_{D-2}) ],  \\ \nonumber
\hat\Omega^2 &=& \frac{R^2}{ \cosh u + \cosh \frac{\tilde \tau}{R} }.
\end{eqnarray}
This metric is  conformal to $\mathbb{R}\times AdS_{d}$.  It is convenient to define the co-ordinate 
\begin{equation}
 \tau = \frac{\tilde \tau}{R}.
 \end{equation}  
The thermal density matrix on this space is given by 
\begin{equation}
\rho_{\cal H} = \frac{e^{2\pi H_{\tau}}  }{ {\cal Z} } , \qquad\qquad {\cal Z}  = {\rm Tr} ( e^{2\pi H_{\tau} } ) .
\end{equation}
where $H_{\tau}$ is the generator of translations along the co-ordinate $\tau$. 
 To evaluate the partition function ${\cal Z}$  using the path integral, 
 we need to identify $ \tau \rightarrow \tau + 2\pi$. Since the causal development of the sphere in Minkowski space  ${\cal D}$ is conformally equivalent 
 to  the hyperbolic cylinder we have the relation
 \begin{equation} \label{relationdh}
 \rho_{\cal D} = U^{-1}  \rho_{\cal H} U =  \frac{ U^{-1} e^{2\pi H_{\tau} }  U }{ {\cal Z} } .
\end{equation}

We now use the relation between $\rho_{\cal D}$ and $\rho_{\cal H}$ to write down 
a formula for the R\'{e}nyi entropy across the sphere $\Sigma$ \cite{Hung:2011nu,Casini:2010kt}.  
The R\'{e}nyi entropy is given by 
\begin{equation}
S_q = \frac{1}{ 1- q} \log {\rm Tr} ( \rho_{\cal D}^q ) .
\end{equation} 
Let us define the free energies 
\begin{equation}
{\cal F}_q = - \log {\cal Z}_q  =   - \log {\rm Tr}  ( \rho_{\cal D}^q) .
\end{equation}
Then using (\ref{relationdh}) we obtain 
\begin{equation} \label{defsq}
S_q = \frac{ q {\cal F}_1 - {\cal F}_q }{ 1-q}.
\end{equation}
Therefore evaluating ${\cal Z}_q$, 
the path integral on the hyperbolic cylinder $S^1 \times AdS_d$ with the identification $\tau \rightarrow \tau + 2\pi q $ enables us to evaluate the R\'{e}nyi entropies 
across $\Sigma$.  The entanglement entropy is given by 
\begin{equation}
S_{EE} = \lim_{q\rightarrow 1} S_q.
\end{equation}
We can also obtain the conformal dimension of the co-dimension two twist operator creating 
the located on the entangling surface using the  free energy on the hyperbolic cylinder. 
This is given by 
\begin{eqnarray}\label{condimtwist}
h_q = \frac{q}{d {\rm Vol }( AdS_d) } 
\left(  \left. \partial_q {\cal F}\right|_{q = 1} - \partial_q {\cal F}  \right) .
\end{eqnarray} 
This formula arises from the fact that $h_q$ essentially captures the energy density due to the 
presence of the defect along the entangling surface \cite{Hung:2014npa,Herzog:2016bhv}. 
The regulated  volume of $AdS_d$ in given in (\ref{voladsd}).

The relationship between $\rho_{\cal D}$ and $\rho_{\cal H}$ is true for conformal field theories. 
For theories which are not conformally invariant, in general there is no such relation. 
The theory of free linearized gravitons in $D=4$ is not conformally invariant.  
Therefore  we should 
a-priori expect no relation between the entanglement entropy across the $3$-sphere 
to the  thermal partition function on the hyperbolic cylinder $S^1 \times AdS_3$. 
One of  the results we would like to show  in this paper is that   the entanglement entropy 
of gravitons across the $3$-sphere evaluated in \cite{Benedetti:2019uej} using canonical quantization
precisely agrees with that evaluated using the partitions functions on the hyperbolic cylinder. 
However the gravitons in this case are massive. The  mass is such that the constant
mode on $S^1$ saturates the 
Brietenholer-Freedmann  bound on $AdS_3$. 

In this next section  we review the case of the conformal scalar in $D$ dimensions
and observe that the corresponding theory on the hyperbolic cylinder saturates the 
Brietenholer-Freedmann bound. 
Then we examine the Maxwell theory in $D=4$ on the hyperbolic cylinder perform the gauge fixing 
and show the  Kaluza-Klein  zero mode of the 
resulting gauge invariant vector on $AdS_3$ saturates its Brietenholer-Freedmann  bound on $AdS_3$.
Motivated by these observations we obtain the  Brietenholer-Freedmann bound 
for symmetric traceless tensors of spin-$s$ in $AdS_d$. 
We  proceed to evaluate the partition function of gravitons on $S^1 \times AdS_3$ .
We show that when the Kaluza-Klein zero mode of the 
gauge invariant  symmetric transverse spin-2 field  on $AdS_3$ has a mass which 
saturates the  Brietenholer-Freedmann bound  then the entanglement entropy evaluated usign
thermal partition functions on the hyperbolic cylinder agrees precisely with that 
evaluated in \cite{Benedetti:2019uej}. 
We  use this observation  and show that the entanglement entropy of 
higher spins in $D=4$ across the  sphere evaluated by thermal partition functions on the 
hyperbolic cylinder using similar considerations also agrees with the conjecture made in \cite{Benedetti:2019uej}.

\subsection{The conformal scalar}

In this subsection, we  review the evaluation of entanglement entropy of the conformally coupled 
massless scalar in even $D \geq 4$ dimensions. 
The Weyl invariant action of the real scalar  in $D$ dimensions is given by 
\begin{equation}
S =- \frac{1}{2}  \int d^{D} x \sqrt{g} (  \partial_\mu\phi \partial^\mu \phi   + \frac{D-2}{4 (D-1)} R \phi^2).
\end{equation}
Note that the mass is due to the non-minimal coupling of the scalar to the metric. 
Consider the metric of the hyperbolic cylinder  $S^1\times AdS_d$ which is given by 
\begin{equation}
ds^2  =  d\tau^2 +  du^2 + \sinh^2  u d\Omega^2_{D-2}.
\end{equation}
Substituting the  Ricci curvature of this space given by  $R = -d( d-1)$. 
 the mass square $m_0^2$  is given by 
\begin{equation}
m_0^2    =    - ( \frac{d-1}{2} )^2 .
\end{equation}
Note that the conformally coupled scalar satisfies the 
Brietenholer-Freedman bound in $AdS_d$, this observation was first made in \cite{Klebanov:2011uf}. 
The partition function is given by 
\begin{equation}\label{freekke1}
{\cal F}_q = - \frac{1}{2} {\rm  Tr} \log [ - \partial_\tau^2 - \Delta_{(0)}   + m_0^2 ] .
\end{equation}
Here $\Delta_{(0)}$ is the spin-0 Laplacian on $AdS_d$ and $\tau$ is identified as 
$\tau \sim \tau + 2\pi q $. 
The eigenfunctions of the spin-0 Laplacian on $AdS_d$ satisfies the condition  
\cite{Camporesi:1994ga}
\begin{equation}\label{eigen}
 \Delta_{(0)} \psi^{\{\lambda, u\}} = -\left[  \lambda^2 + (  \frac{d-1}{2} )^2 \right] 
 \psi_\lambda^{\{\lambda, u\}} , 
 \end{equation}
 where $\{u\}$ refer to other quantum numbers on $AdS_d$. 
Substituting the Fourier decomposition of the field $\phi$ on $S^1$ as well as on $AdS_d$ we obtain
\begin{equation} \label{freekke}
{\cal F}_q = 
\frac{1}{2} \sum_{n = -\infty}^\infty \int_0^\infty d\mu^{(0)}(\lambda)  
\log \left( \lambda^2 + ( \frac{n}{q} )^2  \right).
\end{equation} 
Here $\mu^{(0)}(\lambda)$ is the Plancherel  measure of the real scalar on $AdS_d$. 
For  odd $d\geq 3$, this is given by  \cite{Camporesi:1994ga} \footnote{We adopt the normalizaton 
of the measure given in \cite{Herzog:2016bhv}}. 
\begin{equation}\label{planscalar}
d\mu^{(0)} (\lambda) = \frac { {\rm Vol} ( AdS_d ) }{ 2^{d-1} \pi^{\frac{d}{2} } \Gamma( \frac{d}{2} ) }
\lambda^2 \prod_{j =1}^{ \frac{d-3}{2}} ( j^2 + \lambda^2) .
\end{equation}
where the regulated Volume of $AdS_d$ is given by  \cite{Hung:2011nu}
\begin{equation} \label{voladsd}
{\rm Vol} ( AdS_d )  =  \frac{ 2 \pi^{\frac{d-1}{2} } ( -1)^{ \frac{d+1}{2} } }{ \Gamma( \frac{d+1}{2} ) }
\log( \epsilon) ,
\end{equation}
$\epsilon$ is a dimensionless cut off which regulates the volume of $AdS_d$. 
Comparing the eigenvalues of the spin-0 Laplacian  in $AdS_d$  (\ref{eigen}) and 
the expression in (\ref{freekke1}), shows  that when the mass of the scalar saturates the 
Brietenholer-Freedman bound, the shift in the 
eigen-values of the spin-0 Laplacian  cancels. The constant mode
on $S^1$ has zero eigenvalue at $\lambda =0$ and the system therefore  is gapless. 
The summation over $n$ in (\ref{freekke}) can be done    resulting in 
\begin{equation}
{\cal F}_q = \int_0^\infty  d\mu^{(0)} (\lambda) 
\left [ \log( 1- e^{-2\pi q \lambda} ) + \pi q \lambda \right].
\end{equation}
Substituting this in the expression for R\'{e}nyi entropies  given in (\ref{defsq}) we obtain 
\begin{equation}\label{renyiscald}
S_q = \frac{1}{1- q} \int_0^\infty  d\mu^{(0)} (\lambda)  \left[ q \log( 1- e^{-2\pi  \lambda} ) - 
 \log( 1- e^{-2\pi q \lambda} ) \right].
\end{equation}
As an example, considering
 the Plancherel measure for $D=4, d= 3$ from (\ref{planscalar}) and substituting 
the regulated volume of $AdS_3$  from (\ref{voladsd}) we obtain 
\begin{eqnarray}
S_q = \frac{ ( q+ 1) ( q^2+ 1) }{ 360 q^3} \log ( \epsilon) , \qquad S_{EE} = \frac{1}{90} \log( \epsilon) .
\end{eqnarray}

Since we know the free energy, 
 conformal dimension of the twist operator associated with the entangling surface
 can be evaluated using (\ref{condimtwist})
 This results in 
 \begin{equation}\label{hqval}
h_q = \frac{q}{ 2^{d-1} \pi^{\frac{d}{2} - 1} d \Gamma( \frac{d}{2} ) }
\int_0^\infty  d\lambda \,  (  \lambda^3 \prod_{j =1}^{ \frac{d-3}{2}} ( j^2 + \lambda^2) )  \;
\left( \coth \pi \lambda - \coth \pi q\lambda \right) .
 \end{equation}
 Here we have substituted the Plancherel measure given in (\ref{planscalar}). 
 Evaluating this expression for $D=4, d=3$ we obtain 
 \begin{eqnarray}
 h_q &=& \frac{q}{ 6 \pi } \int_0^\infty d\lambda  \, \lambda^3 
 \left( \coth \pi \lambda - \coth \pi q\lambda \right), \\ \nonumber
 &=& \frac{ q^4- 1}{720 \pi q^3} .
 \end{eqnarray}
 This approach of obtaining the conformal dimension using the partition function on $S^1\times 
 AdS_d$ agrees with that obtained by evaluating the expectation value of the stress 
 tensor on the replica 
 surface using the scalar propagator on the cone \cite{Dowker:1987pk,Dowker:2010bu}. 
 Expressions for the R\'{e}nyi entropies and the conformal dimension $h_q$ for 
   even $D\geq 4$ can be obtained using (\ref{renyiscald}) and (\ref{hqval})
  \footnote{We restrict ourselves to even $D$, since the resulting integrals can be performed
  analytically. However such expressions also exist for odd $D$.}

\subsection{ Free Maxwell field in  ${D=4}$}
\label{maxfield}

In this section we examine the case of the  free $U(1)$ theory on the hyperbolic cylinder. 
This theory has been studied extensively, the logarithmic coefficient of the entanglement 
for the free $U(1)$ theory across the 2-sphere 
 has been extracted using the expectation value of the stress tensor
on de-Sitter space \cite{Dowker:2010bu} and the conformally related hyperbolic cylinder
 \cite{Eling:2013aqa}. 
Both these calculations rely on the gauge invariant two point function of the 
Maxwell field evaluate on the cone or on Rindler space  first done by
\cite{Candelas:1977zza} \footnote{We revisit this
calculation  in appendix \ref{appen2}. }. 
In \cite{Casini:2015dsg} the logarithmic coefficient was extracted  by showing that canonical quantization 
in spherical coordianates
of the Maxwell field results in  2 independent massless scalars  with angular momentum 
$l\geq 1$.  Finally in \cite{Huang:2014pfa}, 
the partition function of the Maxwell field on $S^1\times AdS_3$ was 
evaluated using the heat kernel. 
The above 
methods do not capture the edge modes or the contribution of the electric centre 
\cite{Huang:2014pfa,Donnelly:2015hxa,Donnelly:2014fua,Soni:2016ogt,Moitra:2018lxn}, these 
modes restore the logarithmic coefficient so that it agrees with the trace anomaly. 
However it has been shown recently that these contributions arise only in the presence of
interaction with heavy charges, see \cite{Casini:2019nmu} for a detailed discussion. 
In this paper we focus on the free Maxwell field. 

We  evaluate the partition function of the $U(1)$ theory on the hyperbolic cylinder $S_1\times AdS_3$
by appropriately gauging fixing the theory and then use (\ref{defsq}), (\ref{condimtwist}) 
to evaluate the entanglement entropy/R\'{e}nyi entropies as well as the conformal 
dimension of the twist operator. 
This is a warm up  exercise 
to evaluate the  partition function of gravitons   on $S^1\times AdS_3$. 
The analysis also demonstrates the mass of the  zero mode of the Maxwell field on $S^1$ satisfies the 
Brietenholer-Freedman bound, an observation which was missed earlier and 
which we will use subsequently. 
As mentioned earlier the partition function of the Maxwell theory has been evaluated 
earlier using the heat kernel method in \cite{Huang:2014pfa}. 
However the gauge fixing method adopted here is different 
from that used in that  paper. The method developed here allow us to generalise to to higher spins
as well as anti-symmetric tensors efficiently. 

The Maxwell action is given by 
\begin{equation}
S = \frac{1}{4} \int d^4 x \sqrt{g} (  F_{\mu \nu } F^{\mu \nu}) .
\end{equation}
The action is invariant under the gauge transformation
\begin{equation} \label{gaugetrans}
A_\mu \rightarrow A_\mu + \nabla_\mu \epsilon, 
\end{equation}
where $\epsilon$ is the gauge parameter. 
The background metric is given by 
\begin{equation} \label{ads3s1}
ds^2 = d\tau^2 + du^2 + \sinh^2 u  d\Omega^2_2.
\end{equation}
To fix the gauge we follow the methods developed by 
\cite{Fradkin:1983mq,David:2009xg,Gaberdiel:2010ar}. 
We start by decomposing  the field as 
\begin{equation} \label{chgvar}
A_\mu  = ( A_\tau, \;  A_i^\perp + \nabla_i \phi) , 
\qquad  {\rm where}\qquad  \nabla^i A_i^\perp = 0.
\end{equation}
Here $i$, takes values in the $AdS_3$ direction. 
Under gauge transformation  given in (\ref{gaugetrans}) these fields transform as 
\begin{eqnarray}\label{gaugetrans1}
A_\tau \rightarrow A_\tau + \nabla_\tau \epsilon, \qquad 
A_ i^\perp \rightarrow A_i^\perp, \qquad
\phi  \rightarrow \phi+ \epsilon .
\end{eqnarray}
Substituting  these change of variables, the action becomes 
\begin{eqnarray} \label{newact}
S & =&   \frac{1}{2} \int d^4 x \sqrt{g}  \left[ 
\left( \nabla_i A_j^{\perp} \nabla^i A^{\perp \, j }   -  2 A_i^\perp A^{\perp i }   + 
(  \partial_\tau A_i^\perp)(  \partial_\tau A^{\perp \, i } \right )  \right. 
\\ \nonumber
& &  \qquad \qquad \qquad\qquad 
 +  (  \partial_\tau \nabla _i \phi) (  \partial_\tau \nabla ^i \phi) \\ \nonumber
 & & \left.  \qquad \qquad \qquad\qquad  
 + ( \nabla_i A_\tau )( \nabla^i A_\tau )
-2 \partial_\tau \nabla_i \phi \nabla^i A_\tau   \right].
\end{eqnarray}
Here we have integrated by parts and dropped terms using the condition $\nabla^i A_i^\perp = 0$. 
We have also interchanged the order of derivatives using 
\begin{equation}\label{curvads3}
[ \nabla_i , \nabla_j] A_l =  R^{\;k}_{l\;ij} A_k , \qquad\qquad 
R_{ijkl} = g_{il } g_{jk} - g_{ik} g_{jl}.
\end{equation}
The second line follows due to the fact that $AdS_3$ is a maximally symmetric space with 
unit radius. 
The change of variables in (\ref{chgvar})  also comes with a change in measure which is given by 
(\ref{max-changem})
\begin{equation}\label{chgmea}
{\cal D} A_\mu =   [ {\rm d
et} ( - \Delta_{(0 )} )  ]^{1/2}  {\cal D} A_\tau  {\cal D} A_i^{\perp} {\cal D} \phi, 
\end{equation}
where  $ \Delta_{(0 )}$ is the spin-0 Laplacian on $AdS_3$. 
To treat the zero modes on the $S^1$ separately, 
we write this change of measure into that involving the zero modes on $S^1$ and the rest. 
\begin{eqnarray}
\label{chgmea2}
{\cal D} A_\mu &=&   \left[  ( {\rm d
et} ( - \Delta_{(0 )} )  )^{1/2}  {\cal D} A_\tau  {\cal D} A_i^{\perp} {\cal D} \phi
\right]_{n\neq 0}  \\ \nonumber 
& & \times  \left[ ( {\rm d
et} ( - \Delta_{(0 )} )  )^{1/2}  {\cal D} A_\tau  {\cal D} A_i^{\perp} {\cal D} \phi
\right ]_{n=0}.
\end{eqnarray}

\subsubsection*{Non-zero Kaluza-Klein modes}

Let us first deal with the non-zero modes. 
From the gauge transformation given in (\ref{gaugetrans1}), we see that we can fix the 
gauge $A_\tau =0$. 
This results in the Fadeev-Popov determinant \footnote{To deal with the formal operator
$\partial_\tau$, it is best to think in terms of Fourier modes. Then one just replaces 
$\partial_\tau \rightarrow  in $. }
\begin{equation} \label{fp}
\Delta_{\rm{FP}} = {\rm det} ( \partial_\tau) .
\end{equation}
Therefore we can introduce the Fadeev-Popov determinant along with the 
delta function which chooses this gauge slice
\begin{equation}
{\cal G} = \delta (A_\tau)  {\rm det} ( \partial_\tau) .
\end{equation}
Note that this determinant does not vanish for the non-zero Kaluza-Klein modes. 
Combining all this, the partition function becomes
\begin{eqnarray}
{\cal Z}_{n\neq 0} &=& \int \left[  {\cal D} A_i^\perp \;  (  {\rm d
et} ( - \Delta_{(0 )} )^{1/2}  {\rm det} ( \partial_\tau)   \right]_{n\neq 0} 
\times  \left[  \frac{1}{  {\rm det} ( \partial_\tau) ( {\rm d
et} ( - \Delta_{(0)} ) )^{1/2} }  \right]_{n\neq 0}  \nonumber \\
& &  \times 
\exp \left\{ - \frac{1}{2} \int d^4 x  \sqrt{g}  \left(
 A_{i\, n\neq 0} ^\perp ( -\partial_{\tau}^2  - \Delta_{(1)} - 2 ) 
 A^{\perp i }_{n\neq 0} 
\right) \right\}.
\end{eqnarray}
We have integrated over $A_\tau$ using the delta function constraint. 
The  first two determinants arise from the change in measure  (\ref{chgmea2}) 
and the Fadeev-Popov determinant (\ref{fp}). The determinants in the denominator 
comes from integration over the longitudinal mode
$\phi$. These determinants cancel and we are left with 
the following
\begin{eqnarray} \label{redpart1}
{\cal Z}_{n\neq 0} &=& \int {\cal D}  A_i^\perp  \; 
\exp \left\{ - \frac{1}{2} \int d^4 x  \sqrt{g}  \left(
 A_{i\, n\neq 0} ^\perp ( -\partial_{\tau}^2  - \Delta_{(1)} - 2 ) 
 A^{\perp i }_{n\neq 0} 
\right) \right\}, \\ \nonumber
&=& \left[ \frac{1}{ {\rm det} ( -\partial_{\tau}^2 - \Delta_{(1)} -2)_{n\neq 0}  } \right]^{\frac{1}{2} }.
\end{eqnarray}
Here $\Delta_{(1)}$ is the spin-1 Laplacian on a transverse vector in $AdS_3$.

\subsubsection*{Kaluza-Klein zero modes}

The Kaluza-Klein zero mode demands a separate treatment. 
From (\ref{newact}), we see that the  action for these modes beome
\begin{eqnarray}\label{veczeromod}
S_{n= 0} & =&   \frac{1}{2} \int d^4 x \sqrt{g}  \left[ 
\left( \nabla_i A_j^{\perp} \nabla^i A^{\perp \, j }   -  2 A_i^\perp A^{\perp i }   
 +  \nabla_i A_\tau \nabla^i A_\tau \right)  \right] .
\end{eqnarray}
Here the Kaluza-Klein zero mode of $A_\tau$ 
 along the $S^1$ is gauge invariant, from (\ref{gaugetrans1}) it can be seen 
that it transforms as 
\begin{equation}
A_{\tau, \, n=0} \rightarrow A_{\tau, \, n=0}.
\end{equation}
The action (\ref{veczeromod})  is independent of  the longitudinal mode
$\phi$ and therefore is gauge invariant under 
\begin{equation}
\phi_{n=0}  \rightarrow \phi_{n=0} + \epsilon_{n=0}.
\end{equation}
We fix this gauge by restricting to the slice  
\begin{equation}
{\cal G} = \delta ( \phi_{n=0} ) .
\end{equation}
Note that the Fadeev-Popov
determinant for this choice of gauge slice is unity. 
Using the measure for the zero modes in (\ref{chgmea2}), the functional integral reduces to 
\footnote{We have retained the the integration as $\int d^4 x \sqrt{g}$ which includes the integral over the $\tau$ direction. This just picks up a factor of $2\pi q$ which cancels against the normalisation
of the zero mode. }
\begin{eqnarray} \label{redpart2}
{\cal Z}_{n=0} &=& \int  \left[ {\cal D}A^\perp_{i} {\cal  D}A_\tau  {\rm det} ( -\Delta_{(0)} ) \right]_{n=0} 
\\ \nonumber
& & \times
\exp\left\{ - \frac{1}{2} \int d^4 x  \sqrt{g}  \left(
 A_{i\, n=0} ^\perp (  - \Delta_{(1)} - 2 ) 
 A^{\perp\,  i } _{n=0} 
 + A_{\tau\, n =0} ( - \Delta_{(0)} ) A_{\tau\, n=0}   \right ) \right\} \\ \nonumber
 &=& \left[ \frac{1}{ {\rm det} ( -\Delta_{(1)} -2 ) } \right]^{\frac{1}{2}}.
\end{eqnarray}

It is interesting to note that for the $n\neq 0$ modes, the  Jacobian in the change in 
measure given in (\ref{chgmea2}) is cancelled by the integration of the longitudinal 
field $\phi$. 
However for the $n=0$ modes, the Jacobian is cancelled by the integration of 
$A_\tau$ which is gauge invariant. 
Combining the results from (\ref{redpart1}) and (\ref{redpart2}) 
we can write the  complete partition function as 
\begin{equation} \label{redpart} 
{\cal Z} = \left[ \frac{1}{ {\rm det} ( - \partial_\tau^2 -\Delta_{(1)} -2 ) } \right]^{\frac{1}{2}}.
\end{equation}
The end result involves the determinant of only in the transverse vector on $AdS_3$ including 
all its Kaluza-Klein modes over $S^1$. 
The zero mode of this vector has mass square which is 
given by 
\begin{equation}
m_{s=1}^2 = - 2.
\end{equation}
This mass saturates the Brietenholer-Freedman bound  for a spin-1 field on $AdS_3$. 
We can now Fourier decompose   the vector on $S^1$  and $AdS_3$. 
These observations will be used in section \ref{tenfield} to evaluate the partition function 
of fields of spins $s>1$. 

The eigenvalues of the spin-1 Laplacian on transverse vector on $AdS_3$ is given by
\cite{Camporesi:1994ga}.
\begin{equation}\label{spin1lap}
\Delta_{(1)} \psi_i^{\{\lambda, u \}} = - ( \lambda^ 2 + 2) \psi_i^{\{\lambda, u \}}.
\end{equation}
Substituting  the decomposition of the transverse vector into Fourier modes as well as
the eigenfunctions on $AdS_3$ 
into the partition function (\ref{redpart}),  the free energy is given 
\begin{equation}\label{freeeng1}
{\cal F}_q =  \frac{1}{2} \sum_{n=-\infty}^\infty
\int d\mu^{(1)} \log ( \lambda^2 + \frac{n^2}{q^2}) .
\end{equation}
Note that again that the shift from the eigenvalue of the spin-1 Laplacian  in 
(\ref{spin1lap}) cancels against the mass  in (\ref{redpart}).  Just as in the case of the 
conformal boson, this ensures that the constant mode of the vector on $S^1$ has 
a gapless spectrum as $\lambda$ runs from $0$ to $\infty$. 
The Plancherel measure for the spin-1 field  is given by  \cite{Camporesi:1994ga,Gopakumar:2011qs} 
\begin{equation} \label{spin1mea}
d\mu^{(1)} = \frac{ { \rm Vol} (AdS_3) }{ \pi^2} ( \lambda^2 + 1)  d\lambda.
\end{equation}
In this measure we have incorporated the spin dependent factor $g(s)$  which was determined  by 
\cite{Gopakumar:2011qs} using analytical continuation from the sphere $S^3$. 
For $d=3,  g(s) = 1 $and for $s=0$ and $g(s) =2$ for $s\geq 1$. 
The regularised volume of $AdS_3$ can be read out from (\ref{voladsd}) is given by 
\begin{equation}
{\rm Vol} (AdS_3) = 2\pi \log ( \epsilon)
\end{equation}
Performing the sum over the Kaluza-Klein modes in 
 the free energy (\ref{freeeng1}) and substituting this  into the expression 
for the R\'{e}nyi entropy (\ref{defsq}) we obtain
\begin{eqnarray}
S_q &=& \frac{1}{1-q} \int d\mu^{(1)} \left[
 q \log ( 1- e^{-2\pi \lambda} ) - \log ( 1- e^{ - 2\pi q\lambda} )  \right], \\ \nonumber
 &=& \frac{(q+1)( 31 q^2  +1) }{ 180 q^3}  \log(\epsilon) , \\ \nonumber
 S_1 &=& \frac{16}{45} \log(\epsilon).
\end{eqnarray}
Finally we apply (\ref{condimtwist}) to obtain the conformal dimension of the twist operator localised 
along the entangling surface. 
This results in 
\begin{eqnarray}\label{hqvector}
h_q &=& \frac{q}{3\pi}\int_0^\infty  d\lambda \, 
\lambda (\lambda^2 +1) [ \coth (\pi \lambda) - \coth( \pi q \lambda) ] \\ \nonumber
&=& \frac{(q^2-1)( 11q^2+1) }{ 360 \pi q^3} .
\end{eqnarray}
The above result agrees with  that obtained by the expectation value of the 
stress tensor of the Maxwell field on the replica geometry. 
We revisit this calculation in appendix \ref{appen2}.

One point worth emphasising is that this result for $h_q$ precisely agrees with
the relations found in \cite{Hung:2014npa} between the first derivative and 2nd derivative of 
the $h_q$ with respect to $q$ and the two point and three point functions in 
a CFT.   To verify this let us take the first derivative and the 2nd derivate of the 
expression for $h_q$ given in (\ref{hqvector}), these are given by 
\begin{equation}\label{valderhq}
\partial_q h_q|_{q=1}  = \frac{1}{15\pi}, \qquad  \partial^2_q h_q |_{q=1}  =  -\frac{4}{45\pi}.
\end{equation}
 Let the  parameters determining the $3$ point functions of a CFT by 
given by $a, b, c$ according to \cite{Osborn:1993cr}. 
Then  the relation found in \cite{Hung:2014npa} can be written as 
\begin{equation}\label{expabcrel}
\partial_q h_q |_{q=1} =  \frac{\pi^5 }{180}( 14 a - 2b - 5c) , \qquad
\partial^2_q h_q |_{q=1} = -\frac{\pi^5}{90} ( 16 a - 3b -5c) .
\end{equation}
For the free Maxwell field in $D=4$, the values of $a, b, c$ are given by  \cite{Osborn:1993cr}
\begin{equation}\label{valabc}
a = - \frac{2}{\pi^6}, \qquad b=0, \qquad c= - \frac{8}{\pi^6}.
\end{equation}
Substituting (\ref{valabc}) into (\ref{expabcrel}) we see that it precisely coincides with 
(\ref{valderhq}).

\section{Antisymmetric tensors}\label{antiten}

In this section, we  evaluate the entanglement entropy of   a $2$-form
 in $6$ dimensions across a $4$-sphere. 
This question has been address before. In   \cite{Dowker:2017flz}  the partition function
of this theory on branched spheres which is conformally related to the hyperbolic cylinder was evaluated. 
In \cite{Nian:2015xky} the heat-kernel on   the hyperbolic cylinder of this theory  was assumed to 
factorize  into that on $S^1$ and that of a co-exact $2$-form  on $AdS_5$, this was then used to evaluate 
the R\'{e}nyi entropy.  
Recall, that this theory is conformal and therefore 
using the argument reviewed  in section  \ref{spherhyper},  we can 
evaluate the entanglement entropy by considering the partition function of the 
 anti-symmetric
$2$-form on $S^1 \times AdS_5$. 
Starting from the action
we will fix gauge and reduce the 
partition function to that of a Kaluza-Klein tower of co-exact  $2$-forms on $AdS_5$. 
Thus we prove the factorization which was assumed in \cite{Nian:2015xky}. 
The zero mode of this field  is massless in $AdS_5$ and satisfies the Breitenlohner-Freedman 
bound 
of the $2$-form in $AdS_5$.  
 We also evaluate the conformal dimension $h_q$ of the  corresponding 
co-dimension 2 twist operator. 
We then perform a consistency check by evaluating the conformal dimension using the 
expectation value of the stress tensor on the cone. This  generalises the methods of 
\cite{Candelas:1977zza} to higher rank tensors. 
Finally, we will show that conformal dimensions of the twist operators localized on the 
entangling surface obeys  the relations found in \cite{Hung:2014npa}. 
These relate    the first and 
second derivative of  $h_q$ to the coefficients of the 
three point function of the stress energy tensor of the theory. 
In section \ref{d2form} we generalise the analysis to  $D/2-1$ forms in $D$-dimensions with 
$D=8, 10, 12, 14$.

\subsection{$2$-form  in $D=6$ } \label{sect:2form}

The action of the $2$ form is given by 
\begin{eqnarray}\label{antiact}
S &=&  \frac{1}{12} \int d^6 x \sqrt{g} 
( F_{\mu\nu\rho} F^{\mu\nu\rho} ) , \\ \nonumber
\end{eqnarray}
where the metric is given by 
\begin{equation}
ds^2 = d\tau^2 +  du^2 +  \sinh^2 u d\Omega_{4}^2
\end{equation}
where $\tau$ is identified as $\tau \sim \tau + 2\pi q$, we will label the indices along the 
$AdS_5$ directions with $i, j \cdots$. 
The curvature of $AdS_5$ satisfies
\begin{equation}\label{anticurv}
R_{ijkl} = g_{il} g_{jk} - g_{ik}g_{jl}, \qquad R_{ij} = -4 g_{ij}, \qquad R = - 20. 
\end{equation}
The field strength is defined from the gauge potential $A_{\mu\nu}$  by  
\begin{equation}
F_{\mu\nu\rho} = \nabla_{\mu} A_{\nu\rho} + 
\nabla_{\nu} A_{\rho\mu} + 
\nabla_{\rho} A_{\mu\nu}. 
\end{equation}
The action (\ref{antiact})  is invariant under the gauge transformation
\begin{equation}\label{antigt}
\delta A_{\mu \nu} = 
\nabla_{\mu}\epsilon_{\nu}^\perp - \nabla_{\nu} \epsilon_{\mu}^\perp, 
\qquad  \nabla^\mu \epsilon_\mu^\perp = 0
\end{equation} 
Note that the longitudinal component of $\epsilon^\mu$  which we 
denote by $\tilde\epsilon$ does not enter the gauge transformation. 
This point will be important subsequently.

Following our pervious analysis, we start by decomposing the field as 
\begin{eqnarray} \label{antichgv}
& & A_{\mu\nu} = \{ A_{\tau i }, A_{ij} \},  \qquad\qquad A_{\tau i} = A_{\tau i}^\perp + \nabla_i \chi_\tau
\qquad \nabla^i A_{\tau i }^\perp =0, \\ \nonumber
& &A_{ij} = A_{ij}^\perp + \nabla_i \chi_j ^\perp - \nabla_j \chi_i ^\perp, \qquad\qquad
\nabla^i A_{ij}^\perp = 0, \qquad \nabla^i \chi_i^\perp = 0.
\end{eqnarray}
Note that $A_{ij}^\perp$ is a co-closed 2-form. 
In terms of these variables the action becomes
\begin{eqnarray}\label{antiact2}
S &=& \int d^6 x \sqrt{g} ( {\cal L}_1 + {\cal L}_2 + {\cal L}_3 ) , \\ \nonumber
{\cal L}_1  &=&  \frac{1}{4} \left[ 
( \nabla_i A_{ \tau j }^\perp - \nabla_j A_{ \tau i }^\perp )
- 2\partial_\tau(  A_{ij} ^\perp + \nabla_i \chi_j^\perp - \nabla_j \chi_i^\perp )   \right]
 ( \nabla^i A^{\tau j\, \perp} - \nabla^j A^{  \tau i \, \perp} ), 
 \\ \nonumber
 {\cal L}_2 &=& \frac{1}{2} \chi^{i \, \perp} \partial_\tau^2 (  \Delta_{(1)}  + 4 ) \chi_i^\perp, 
 \\ \nonumber
 {\cal L }_3 &=& \frac{1}{4} A_{ij}^{\perp} (-  \Delta_{{\rm HdR}} - \partial_\tau^2 ) A^{ij \, ^\perp}, 
\end{eqnarray}
where $\Delta_{\rm HdR}$ refers to the  Hodge-de Rham Laplacian, which acts  on 
co-closed  $2$-forms as 
\begin{equation}
 \Delta_{{\rm HdR}} A^{ij \perp} = \nabla_k ( \nabla^k A^{ ij \, \perp } + 
 \nabla^i A^{ jk  \, \perp }  + \nabla^j A^{ k i  \, \perp }  ), 
 \end{equation}
 $\Delta_{(1)}$ is the Laplacian acting on transverse vectors on $AdS_5$.  To obtain the action 
 in (\ref{antiact2}) we have interchanged covariant derivatives at the 
 cost of introducing curvature terms and then  substituted for the curvature as given in 
 (\ref{anticurv}). 
 We also decompose the gauge parameter as
 \begin{equation}
 \epsilon_i = \epsilon_i^\perp + \nabla_i \epsilon.
 \end{equation}
 The gauge transformation (\ref{antigt})   becomes
 \begin{eqnarray}\label{antigt2}
& & \delta A_{\tau i } ^\perp   = \partial_\tau \epsilon_i^\perp, \qquad 
\delta \chi_\tau = \partial_\tau \epsilon - \epsilon_\tau, \\ \nonumber
&& \delta A_{ij}^\perp =0, \qquad \delta \chi_i^\perp = \epsilon_i^\perp.
 \end{eqnarray}
 The gauge parameters $\epsilon_\tau, \epsilon$ are related by 
 \begin{equation}\label{anticonstraint}
 \partial_\tau \epsilon _\tau + \Delta_{(0)}  \epsilon = 0
 \end{equation}
 This results from the  transversality condition (\ref{antigt}) on the gauge parameter $\epsilon_\mu$. 
 The  Jacobian for the change of variables  (\ref{antichgv}) is given by  (\ref{2form-changem1}), 
 ( \ref{2form-changem2}), 
 \begin{eqnarray}
 {\cal D} A_{\mu\nu} = { \cal D} A_{ \tau i }^\perp 
 {\cal D} \chi_\tau {\cal D}A_{ij}^\perp {\cal D} \chi_i^\perp  \times J , 
 \\ \nonumber
 J  =  \left[ {\rm det}\left(  ( -\Delta_{(1)} - 4) ( -\Delta_{(0)} ) \right)  \right]^{\frac{1}{2} }.
 \end{eqnarray}
 where $\Delta_{(1)}, \Delta_{(0)}, $ is the spin-1 and spin-0 Laplacians on $AdS_5$. 
 We can split this measure  into that 
 for the non-zero Kaluza-Klein modes on $S^1$  and zero Kaluza-Klein modes. 
 \begin{eqnarray}\label{antiJaco}
  {\cal D} A_{\mu\nu} &=& \left[ { \cal D} A_{ \tau i }^\perp {\cal D} \chi_\tau
  {\cal D} A_{ij}^\perp {\cal D} \chi_i^\perp  \times J
  \right]_{n\neq 0}, \\ \nonumber
  &&\times   \left[ { \cal D} A_{ \tau i }^\perp  {\cal D} \chi_\tau
 {\cal D} A_{ij}^\perp {\cal D} \chi_i^\perp  \times J
  \right]_{n=  0}.
 \end{eqnarray}

 \subsubsection*{Non-zero Kaluza-Klein modes}
 
 The gauge symmetry in (\ref{antigt2}) allows us to fix the gauge $A_{\tau i }^\perp, \chi_\tau=0$. 
 Let us see this more explicitly,  for non-zero Kaluza-Klein modes the gauge transformation 
 on the field $A_{\tau i }$ becomes
 \begin{equation}
 \delta A_{\tau i }^\perp = \partial_\tau \epsilon_i^\perp, 
 \qquad \delta \chi_\tau = \frac{1}{\partial_\tau } ( \partial_\tau^2 + \Delta_{(0)} ) \epsilon.
 \end{equation}
 To  obtain the second equation we have used the constraint (\ref{anticonstraint}) 
 relating $\epsilon_\tau$ and $\epsilon$. 
  It is clear choosing $\epsilon_i^\perp, \epsilon$ we can gauge away $A_{\tau i}$ for the 
  non-zero modes. 
 Therefore we can impose this condition using  a delta function along with the corresponding
 Fadeev-Popov determinant in the path integral
 \begin{equation} \label{antiFP}
 {\cal G} = \delta ( A_{ i \tau } ^\perp) \delta(\chi_\tau)  \left[ {\rm det} \left( ( \partial_\tau )^5  
 \left( \frac{1}{\partial_\tau }\right) ( \partial_\tau^2 + \Delta_{(0)} )  \right)  \right]_{n\neq 0} .
 \end{equation}
 Note that since the  delta function is in the co-ordinate system in which the gauge parameters
 are $\epsilon_i^\perp$ and $\epsilon, \tilde\epsilon$, we need to transform the integration over 
 gauge group from $\epsilon_\mu$.  This is then divided out. 
 The change in measure is given in  (\ref{gauge-changem1}), 
 \begin{eqnarray}\label{antij}
 &&{\cal D} \epsilon_i^\perp {\cal D} \epsilon {\cal D} \tilde\epsilon =  {\cal D} \epsilon_\mu \times   J', \\ \nonumber
 && J' = \left[
 {\rm det} \left( (  \partial_\tau^2 + \Delta_{(0)} ) \left( \frac{1}{\partial_\tau }\right)  \right)_{n\neq 0}  \right]^{-1} [ {\rm det} ( - \Delta_{(0)} ) ]^{-\frac{1}{2}}.
 \end{eqnarray}

 Integrating over $A_{\tau i }^\perp, \chi_\tau$  using 
  the delta function we obtain the following action
 \begin{equation}
 S =  \int d^6 x \sqrt{g}  ( {\cal L}_2 + {\cal L}_3) .
 \end{equation}
 After performing the functional integrals over $A_{ij}^\perp$  and $\chi_i^\perp$ we obtain
 the partition function 
 \begin{equation}
{\cal Z}_{A_{ij}^\perp, \chi_i^\perp} = 
\left[ 
{\rm det} \left(  ( - \partial_\tau^2  - \Delta_{HdR} )  ( - \Delta_{(1)} - 4) (  \partial_\tau^2)^5 \right) 
\right]^{\frac{1}{2} }_{n\neq 0}.
\end{equation}
The complete partition function is obtained by multiplying  the one loop determinants along with the
Fadeev-Popov determinant in (\ref{antiFP}),   determinant of the Jacobian  $J$ in (\ref{antiJaco}) as 
well as $J'$ in (\ref{antij}). 
This results in 
\begin{equation}\label{antipart1}
{\cal Z}_{n\neq 0} = 
\left[ \frac{1}{ {\rm det}  ( - \partial_\tau^2  - \Delta_{{\rm HdR}} )_{n\neq 0 } }\right]^{-\frac{1}{2}}.
\end{equation}
Note that, just as in the case of $U(1)$ field and the graviton, we observe that the 
determinants involving the change of measure as well as the Fadeev-Popov determinant 
cancels  in the final expression for the partition function.

\subsubsection*{Kaluza-Klein zero mode}

The action in the Kaluza-Klein zero mode sector reduces to 
\begin{eqnarray}
S  &=& \int d^6 x \sqrt{g} ( {\cal L }_1' + {\cal L}_3 ) , \\ \nonumber
{\cal L}_1' &=& \frac{1}{4} 
( \nabla_i A_{\tau j}^\perp - \nabla_j A_{ \tau i }^\perp )
( \nabla^i A^{\tau j \, \perp} - \nabla^j A^{\tau i \, \perp} ), \\
\nonumber
{\cal L}_3 &=& \frac{1}{4} A_{ij}^{\perp} (-  \Delta_{{\rm HdR}}  ) A^{ij \, ^\perp}.
\end{eqnarray}
The action admits the gauge symmetry
\begin{eqnarray}
\delta A_{\tau i }^\perp  = 0, \qquad \delta \chi_\tau =  - \epsilon_\tau, \qquad
\delta A_{ij}^\perp =0, \qquad \delta \chi_i^\perp  = \epsilon_i^\perp. 
\end{eqnarray}
We can fix the gauge by choosing the slice 
\begin{equation}\label{antigaugslicez}
{\cal G} = \delta( \chi_\tau) \delta ( \chi_i^\perp).
\end{equation}
Since this slice is in the frame where the gauge group is parameterized in  $\epsilon_i^\perp, \epsilon,\epsilon_\tau$ we need to transform it to the parameters $\epsilon_\mu$. It is the  volume of 
gauge group $\int {\cal D} \epsilon_\mu$ which is divided out. 
Note that for the zero modes, the condition $\nabla^\mu \epsilon_\mu =0$ in (\ref{anticonstraint}) reduces
to $\nabla^i\epsilon_i =0$. 
The change in measure is given by (\ref{gauge-changem2}). 
\begin{equation}\label{antiggchg}
{\cal D} \epsilon_\tau {\cal D} \epsilon_i^\perp{\cal D} \epsilon = {\cal D} \epsilon_\mu J', \qquad
J' = [ {\rm det}( - \Delta_{(0)} ) ]^{-\frac{1}{2}}_{n=0 }.
\end{equation}

We can now perform the integrations over $\chi_\tau, \chi_i^\perp$ using the delta function and 
then perform the integral over the gauge invariant variables $A_{\tau i}^\perp,  A_{ij}^\perp$. 
This results in the following one loop determinant. 
\begin{eqnarray}
{\cal Z}_{A_{\tau i}^\perp,  A_{ij}^\perp} = 
\left[ {\rm det} \left( ( -\Delta_{HdR} )  ( -\Delta_{(0)} - 4) \right) \right]^{-\frac{1}{2}}.
\end{eqnarray}
Combining the Jacobians from the change of measure in (\ref{antiJaco}) and that for the 
gauge group in (\ref{antiggchg}) we obtain the following partition function
\begin{equation}\label{antipart2}
{\cal Z} = \left[ \frac{1}{ {\rm det}  ( -\Delta_{HdR} ) _{n =0} } \right]^{\frac{1}{2}}.
\end{equation}
Again we see a similar phenomenon, the one loop determinant of $A_{\tau i}^\perp$
cancels the change in measure in the zero mode sector. 

Combining the partition function for the both the non-zero Kaluza-Klein modes 
from (\ref{antipart1}) and (\ref{antipart2}) and the 
zero modes we obtain 
\begin{equation}
{\cal Z} = 
\left[ \frac{1}{ {\rm det}  ( - \partial_\tau^2  - \Delta_{{\rm HdR}} ) }\right]^{-\frac{1}{2}}.
\end{equation}
In the end, the final result is again simple. It involves the one loop determinant of all 
Kaluza-Klein
the co-closed forms  on $AdS_5$.   This was the assumption used by the authors of 
\cite{Nian:2015xky} to evaluate of the R\'{e}nyi entropy of the self dual 2-form. 

The eigenvalue of the Hodge-deRham Laplacian on $AdS_d$ on co-closed a $p$-form $u$  is given by 
\begin{equation} \label{eigenhdr}
\Delta_{{\rm HdR}} \psi^{\{ \lambda, u\}}_{i_1i_2  \cdots i_p}= 
-\left[ \lambda^2  \right]
 \psi^{\{ \lambda, u\}}_{i_1i_2 \cdots i_p}.
\end{equation}
The eigenvalues do not vanish and therefore they are  co-exact forms. 
The Plancherel measure for co-exact $p$-forms  on $AdS_d$
for odd $d$  is given by  \cite{CAMPORESI199457} 
\begin{eqnarray}
d\mu_{p-{\rm form}} (\lambda) &=& \frac{\hat g(p) {\rm Vol} (AdS_d )}
{ 2^{d-1} \pi^{\frac{d}{2} } \Gamma( \frac{d}{2} ) } \frac{1}{\lambda^2 + ( \frac{d-1}{2} - p)^2 } \prod_{ j =0}^{\frac{d-1}{2}  } ( \lambda^2 + j^2) , 
\\ \nonumber
\hat g(p) &=& \frac{ ( d-1)!}{ p! ( d- p-1) !}.
\end{eqnarray}
As a simple check, note that for $d=3, p=1$ the measure reduces to that of the vector  given in 
(\ref{spin1mea}). 

Let us now apply these results to the case of the $2$-form on $S^1\times AdS_5$. 
We can use the decomposition of the co-closed form into eigenfunctions on $AdS_5$ 
 and the Kaluza-Klein
modes on $S^1$ to write the free energy as
\begin{eqnarray}
{\cal F} = \frac{1}{2} \sum_{n} \int d\mu_{2-{\rm form} } ( \lambda) \log\left( \lambda^2 + \frac{n^2}{q^2}
\right). 
\end{eqnarray}
Here we have used the expression for the eigenvalues of the Hodge deRham Laplacian 
given in (\ref{eigenhdr}). 
The Plancherel measure is given by 
\begin{equation} 
d\mu_{2-{\rm form}} (\lambda) = \frac{1}{2{\pi^3} } {\rm Vol} (AdS_5)  ( \lambda^2 +1) ( \lambda^2 + 4). 
\end{equation}
Before going ahead, 
let us examine if the  constant mode  of the $2$ form along the $S^1$ satisfies the  Breitenlohner-Freedman  bound in $AdS_5$. 
Given a massive $p$-form in $AdS_d$, the Breitenlohner-Freedman  bound is given by 
\cite{Witten:1998qj} \footnote{We can use the mass-conformal dimension of a $p$-form 
in \cite{Witten:1998qj}, $ \Delta = \frac{d-1}{2} + \sqrt{   m^2 + ( p - \frac{d-1}{2} )^2} $   to read out 
Breitenlohner-Freedman  bound .}
\begin{equation}
m^2 \geq  - ( p - \frac{d-1}{2} )^2 .
\end{equation}
Therefore a massless $\frac{d-1}{2}$ from in $AdS_d$ always saturates the Breitenlohner-Freedman bound. 
Note that the constant mode along the $S^1$ of the $2$-form  massless in $AdS_5$. 
The equations of motion it satisfies is $\Delta_{HdR} A_{ij}^\perp =0$, the masses arises
only from the curvature couplings. 
Therefore the constant mode of the  conformal $2$-form in $S^1\times AdS_5$ saturates the 
Breitenlohner-Freedman bound. 
Now using  (\ref{defsq}) we are led to the following integral 
for the R\'{e}nyi entropy
\begin{eqnarray}\label{2formre}
S_q &=&\frac{1}{1-q} \int d\mu_{2-{\rm form}} (\lambda) \left[ q \log( 1- e^{-2\pi  \lambda} ) - 
 \log( 1- e^{-2\pi q \lambda} ) \right], \\ \nonumber
&=& - \frac{ ( q+1) ( 877 q^4  + 37 q^2 + 2) }{ 5040 q^5} \log \epsilon.
\end{eqnarray} 
The entanglement entropy is given by 
\begin{equation}\label{2forme2}
S_1 = -\frac{229}{630} \log \epsilon.
\end{equation}
Similarly we use (\ref{condimtwist}) to 
evaluate  the conformal dimension of the twist operator corresponding to the 
spherical entangling surface. We obtain
\begin{eqnarray}\label{2formhq}
h_q &=& \frac{q}{ 10\pi^2 } \int_0^\infty d\lambda  \lambda ( \lambda^2 + 1) ( \lambda^2 + 4) 
( \coth \pi \lambda -\coth\pi q \lambda) ,\\ \nonumber
&=& \frac{ ( q^2-1) ( 191 q^4 +  23 q^2  +2 ) }{ 5040 \pi^2 q^5}.
\end{eqnarray}

The result in (\ref{2formre}), (\ref{2forme2}), 
(\ref{2formhq}).  
 misses the  contribution of the edge modes at the boundary of the entangling surface, 
that shift the  result so that it agrees with the conformal anomaly. 
Perhaps just as in the discusion for the $U(1)$ theory  in \cite{Casini:2019nmu},  the evaluation of the 
entanglement entropy  using the partition function  on $S^1\times AdS_5$ is the result 
for the free $2$-form theory and not for that in presence of $2$-form charges. 
We  proceed to evaluate $h_q$ using the expectation value of the 
stress tensor on the conical geometry and show that it precisely coincides with the above
value. 
This provides us a check of our path integral methods in particular the gauge fixing procedure 
we adopted.

\subsubsection*{Twist operator weights  from  the replica  cone}

In a conformal field theory, 
the conformal dimension associated with the twist operator  localized on the spherical 
entangling surface  is the same as that of the twist operator  $\sigma_q$ creating a  co-dimension 2
planar defect or a planar entangling surface since these surfaces can be conformally mapped 
to each other.   By conformal invariance the expectation value of the stress tensor on a planar 
defect completely determined  by  the dimension of the twist operator. 
To be specific let us suppose at the time slice $x^1=0$, the planar entangling surface is at $x^2=0$. 
We label the coordinates $a, b$.  
let all the rest of the longitudinal directions by labeled by $x^i, i =3, \cdots D$ and 
$r = \sqrt{  (x^1)^2 + (x^2)^2}$ be the transverse distance from the defect. 
Then the expectation value of the stress tensor takes the form
\begin{eqnarray}\label{stressdefhq}
&&\langle T_{ij} \sigma_q \rangle  = - \frac{h_q}{2\pi} \frac{\delta_{ij}}{ r^D }, 
\qquad \langle  T_{i a} \sigma_q \rangle   =0, \\
\nonumber
&&\langle T_{ab} \sigma_q \rangle  = \frac{h_q}{2\pi} \frac{ (D-1) \delta_{ab} - D n_a n_b }{ r^D}, 
\qquad n^a = \frac{ x^a}{r }.
\end{eqnarray}

To evaluate these expectation values we can adopt the replica trick. 
We consider the path integral on $q$ copies of the CFT is glued along the cut $x^2>0$. 
When one uses polar coordinates  $r, \theta$ to describe the transverse directions 
$x^1, x^2$, the CFT is then on a cone, in which $\theta$ is identified with the periodicity 
$\theta \sim \theta + 2\pi q$. 
We can then evaluate the expectation values of the stress tensor using propagators 
on this cone by the  point split method.  We adopt this procedure for the 
 stress tensor of the $2$ form in $D=6$.\footnote{Appendix revisits this  calculation for the  $U(1)$ theory 
in $D=4$ .}. 
 
The stress tensor can be derived from the action given in (\ref{antiact}). 
It is given by 
\begin{eqnarray}\label{2formstress}
T_{\mu \nu} &=& \frac{1}{2} F_{\mu\rho\sigma} F^{\, \rho\sigma}_{\nu} - \frac{\delta_{\mu\nu}}{12}
F_{\rho\sigma\delta} F^{\rho\sigma\delta}, \\ \nonumber
F_{\mu\nu\rho} &=& \partial_\mu A_{\nu\rho} + \partial_{\nu} A_{\rho\mu} + \partial_{\rho} A_{\mu\nu}.
\end{eqnarray}
We should first fix gauge and then write down the propagator of the 2-form consistent 
with this gauge. 
The $2$ form admits the gauge transformation
\begin{equation}
A_{\mu\nu}\rightarrow A_{\mu\nu} + \partial_\mu \epsilon_\nu^\perp  - \partial_\nu \epsilon_\mu^\perp
\end{equation}
Note that the longitudinal component of the parameter $\epsilon_\mu$ does not contribute to the 
gauge transformation. 
Using this symmetry we can fix the gauge 
\begin{equation} \label{trgcond}
\partial ^\mu A_{\mu\nu} =0.
\end{equation}
Therefore the equation of motion is given by 
\begin{equation}\label{eom2form}
\Box A_{\mu\nu} =0.
\end{equation}
This still allows us the gauge transformation  
\begin{equation}
A_{\mu\nu}' = A_{\mu\nu} + \partial_\mu \epsilon_\nu^\perp  - \partial_\nu \epsilon_\mu^\perp, 
\qquad {\rm with}, \quad \Box \epsilon_\mu^\perp =0.
\end{equation}
Then given a 2-form which satisfies the gauge condition (\ref{trgcond}) and (\ref{eom2form}) we can make a further gauge transformation so that the 2-form in the new frame satisfies the condition
\begin{equation}\label{newgcond}
\partial^a A_{a \mu}^\prime=0,  \qquad \partial^i A_{i \mu} ^\prime = 0   
 \qquad  a\in\{ 1,2\}, \; i \in \{ 3, 4, 5, 6 \} .
\end{equation}
To do this, choose the gauge parameter as
 \footnote{This equation  and similar subsequent equations seem formal.  It can be concretely 
 understood 
in terms of the Fourier space in the $3, 4, 5,6$ directions. The theory  has translational symmetry in these directions. }
\begin{equation}
\epsilon_\mu^\perp  = - \frac{ \partial^i A_{i \mu }}{ \widetilde \nabla^2}, \qquad\qquad \widehat\nabla^2 = \sum_{i=3}^6\partial_i^2.
\end{equation}
It is easy to see that using this  additional gauge transformation the  $2$-form  satisfies 
the more restrictive 
condition (\ref{newgcond}). The gauge parameter also satisfies $\Box\epsilon_\mu^\perp =0$.

Since we need to work on the replica cone it is convenient to 
use the polar coordinates 
\begin{equation}
ds^2 = dr^2 + r^2 d\theta^2 + \sum_{i=3}^6 dx^i dx^i , \qquad \theta \sim \theta +2\pi q.
\end{equation}
Let us define
\begin{equation}
\hat G_{\mu\nu}^{\;\;\mu'\nu'} (x, x') = \langle A_{\mu\nu} ( x)  A^{\mu'\nu'} (x') \rangle .
\end{equation}
Then the Green's function  of the $2$ form satisfying the gauge condition (\ref{newgcond}) is 
given by 
\begin{eqnarray} \label{gauginvp1}
& & \tilde G_{ab}^{\;\; a'b'} (x, x') = \tilde G_{ab}^{\;\; ij} (x, x') =\tilde  G_{ij}^{\;\; a'b'} (x, x') =0 , \\ \nonumber
& & \tilde G_{ai}^{\;\; b' j'} ( x, x') = -\left[ (  \frac{ - P_a P^{b\prime}}{\widehat \nabla^2} ) 
( \delta_i^j - \frac{ \partial _i \partial^j  }{ \widehat\nabla^2 } )  \right] \tilde G (x, x') ,  
\qquad \widehat\nabla^2 = \sum_{i=3}^6 \partial_i ^2 , \\ \nonumber
& &  \tilde G_{ai}^{\;\; b' j'} ( x, x')  = -  \tilde G_{ia}^{\;\; b' j'} ( x, x')  = - \tilde G_{ai}^{\;\; j' b'} ( x, x') =
\tilde G_{ia}^{\;\; j' b'} ( x, x') ,  \\ \nonumber
& & \tilde G_{ij}^{\;\; k'l'} ( x, x') =- \left[
( \delta_{ik} - \frac{\partial_i \partial^k}{ \widehat \nabla^2 } ) 
( \delta_{jl} - \frac{\partial_j\partial^l}{ \widehat \nabla^2 } ) 
- ( \delta_{il} - \frac{\partial_i \partial^l}{ \widehat\nabla^2 } ) 
( \delta_{jk} - \frac{\partial_j\partial^k}{ \widehat \nabla^2 } )  \right ] \tilde G(x, x').
\end{eqnarray}
where $\tilde G(x, x')$ is the Green's function of a massless scalar on the cone.
This can be found in many works, see  \cite{Nozaki:2014hna} for a recent reference. 
\begin{eqnarray}\label{gauginvp}
\tilde G ( x, x') &=&  \frac{f_q( a, \theta, \theta') + g_q( a, \theta, \theta')}{
4\pi^3 q^2 ( rr')^2 ( a - a^{-1)} )^2}, \\ \nonumber
\frac{a}{ 1+a^2} &=&   \frac{rr'}{ ( x-x')^i ( x-x')_i + r^2 + r^{\prime 2}}, \\ \nonumber
f_q(a, \theta, \theta') &=&  
\frac{2 \left(  ( a^\frac{1}{q} + a^{-\frac{1}{q}}) \cos \left( \frac{\theta - \theta'}{q} \right) - 2 \right) }{
\left( a^\frac{1}{q} + a^{-\frac{1}{q}} - 2 \cos \left( \frac{\theta - \theta'}{q} \right)  \right)^2 }, 
\\ \nonumber
g_q( a, \theta, \theta') &=& \frac{  ( a + a^{-1} ) ( a^{\frac{1}{q}} - a^{-\frac{1}{q} } ) } {\left( 
a^{\frac{1}{q}} +  a^{-\frac{1}{q} } - 2 \cos \left( \frac{\theta - \theta'}{q} \right)  \right) ( a - a^{-1} ) }.
\end{eqnarray}
and $P_a$ are defined as
\begin{eqnarray}\label{defpa}
P_a = \epsilon_{a b} g^{bc} \nabla_c , \qquad\qquad
\epsilon_{12} = -\epsilon_{21} = r,  \epsilon_{11} = \epsilon_{22} =0.
\end{eqnarray}
Note that the first line of (\ref{gauginvp1}) follows from the gauge condition 
$\nabla^a \tilde G_{a  \mu}^{\;\rho'\sigma'}  =0$ 
together with the anti-symmetry property of the the Green's function. 

To evaluate the expectation value of the stress tensor on the replica cone, we take the 
expression of the stress tensor given in (\ref{2formstress}) 
and separate the two points ,and then take the expectation values, using the Green's function 
in (\ref{gauginvp1}) 
finally take the coincident limit.  Therefore we have 
\begin{equation}
\langle T_{\mu\nu} \sigma_q \rangle  =  \lim_{x\rightarrow x'}
\frac{1}{2} \langle F_{\mu\rho\sigma} (x) F^{\, \rho\sigma}_{\nu} (x')  \rangle - \frac{\delta_{\mu\nu}}{12}
\langle F_{\rho\sigma\delta} (x)  F^{\rho\sigma\delta}(x') \rangle .
\end{equation}
On taking the coincident limit there are short distance  divergences which are identical to 
that seen on using the propagator with $q=1$. 
To regulate these divergence we replace the  propagator by the regulated propagator given by 
\begin{equation}
 G_{\mu\nu}^{\;\;\mu'\nu'} (x, x') =   \tilde G_{\mu\nu}^{\;\;\mu'\nu'} (x, x')|_{q} -   \tilde G_{\mu\nu}^{\;\;\mu'\nu'} (x, x')|_{q=1}.
 \end{equation}
For example performing  these steps  on the first term in the stress tensor we obtain 
\begin{eqnarray}
  & &\langle  F_{a\mu\nu}(x)  F^{b'\mu'\nu'} (x')  \rangle 
        =(\nabla_{a}\nabla^{b'}G_{\mu\nu}^{\;\;\mu'\nu'}+\nabla_{a}\nabla^{\mu'}G_{\mu\nu}^{\;\;\nu'b'}+\nabla_{a}\nabla^{\nu'}G_{\mu\nu}^{\;\;b'\mu'})\nonumber\\
                                              & & +(\nabla_{\mu}\nabla^{b'}G_{\nu a}^{\;\;\mu'\nu'}+\nabla_{\mu}\nabla^{\mu'}G_{\nu a}^{\;\;\nu' b'}+\nabla_{\mu}\nabla^{\nu'}G_{\nu}^{\;\;b' \mu'}) \\ \nonumber
                                              &&
                                               +(\nabla_{\nu}\nabla^{b'}G_{a\mu}^{\;\;\mu'\nu'}+\nabla_{\nu}\nabla^{\mu'}G_{a\mu}^{\;\;\nu'b'}+\nabla_{\nu}\nabla^{\nu'}G_{a\mu}^{\;\;b'\mu'})\nonumber\\
                                         && =\Big[-6 \nabla_{a}\nabla^{b'} +6 \frac{(\nabla_a P_c- \nabla_c P_a)}{\widehat\nabla^2}\frac{(\nabla^{c'}P^{b'}-\nabla^{b'}P^{c'})}{\widehat \nabla^2}+6 P_{a} P^{b'}\Big] 
                                         G(x, x' ). \nonumber
                                        \end{eqnarray}
To arrive at the last line in the above equation, we have used similar manipulations and 
several identities listed  in 
appendix \ref{appen2}.  We define
\begin{equation}
G(x, x') = \tilde G( x, x')|_{q} -   \tilde G( x, x')|_{q=1} .
\end{equation}
Choosing specific values for the $a, b$ indices, the result  for the non-trivial components 
along the transverse direction of the defect are given by 
\begin{eqnarray}\label{ff1}
\langle F_{\theta\mu\nu}F^{\theta' \mu'\nu'} \rangle  &=& \left[-\frac{6}{r^2}
\frac{\partial^2}{\partial\theta\partial\theta'} 
+6 \frac{\partial^2}{\partial r\partial r'} 
-6\widehat\nabla^2\right]G (x, x')  , \\ \nonumber
\langle F_{r\mu\nu}F^{r' \mu'\nu'}  \rangle  &=& \left[\frac{6}{r^2}\frac{\partial^2}{\partial\theta\partial\theta'} 
-6  \frac{\partial^2}{\partial r\partial r'} -
6\widehat\nabla^2\right] G(x, x') .
\end{eqnarray}
Similarly  going through the analysis  to evaluate the non-zero components along the transverse
directions we obtain
\begin{eqnarray}\label{ff2}
\langle F_{i\mu\nu}F^{j'\alpha'\beta'}  \rangle = 3 \delta_i^j \widehat\nabla^2 G( x, x') .
\end{eqnarray}
Combining (\ref{ff1}) and (\ref{ff2} we get
\begin{equation} \label{ff3}
\langle F_{\mu\nu\rho}F^{\mu'\nu'\rho'} \rangle = 0.
\end{equation}
Therefore the components of the stress tensor are given by 
\begin{eqnarray}\label{tstress1}
\langle T_{\theta}^\theta   \sigma_q \rangle &=&  \lim_{x\rightarrow x'}  \frac{1}{2}  \langle F_{\theta\mu\nu}F^{\theta' \mu'\nu'} \rangle  , \\ \nonumber
&=&  \lim_{x\rightarrow x'}  \left[-\frac{3}{r^2}\frac{\partial^2}{\partial\theta\partial\theta'}
+3\frac{\partial^2}{\partial r\partial r'} -3\widehat\nabla^2\right]G (x, x'), \\ \nonumber
&=& \frac{ ( q^2 - 1) ( 191 q^4 + 23 q^2 +2)} { 1008\pi^2 q^6 r^6}.
\end{eqnarray}
Proceeding along the same lines we obtain
\begin{eqnarray}\label{tstress2}
\langle T_{r}^r   \sigma_q \rangle &=&\lim_{x\rightarrow x'}  \left[\frac{3}{r^2}\frac{\partial^2}{\partial\theta\partial\theta'}
-3\frac{\partial^2}{\partial r\partial r'}
-3\widehat\nabla^2\right] G(x, x') , \\ \nonumber
&=& \frac{ -191 q^6 + 168 q^4 + 21 q^2 + 2}{ 5040\pi^2 q^6 r^6}, \\ \nonumber
\langle T_{i}^i   \sigma_q \rangle &=&\lim_{x\rightarrow x'} 
 \frac{3}{2} \delta^j_i \widehat\nabla^2 G( x, x') , \\ \nonumber
&=& - \frac{( q^2 -1) ( 191 q^4 + 23 q^2 + 2) }{ 5040\pi^2 q^6 r^6} .
\end{eqnarray}

To evaluate $h_q$  and also to confirm the form for the expectation value of the stress tensor given in  (\ref{stressdefhq}), we convert back to cartesian coordinates by 
setting $ x^1 = r \cos\theta, x^2 = r\sin\theta$ and this transformation on the stress tensors in 
(\ref{tstress1}) and (\ref{tstress2}) to obtain 
\begin{eqnarray}
\langle T_{11} \sigma_q\rangle &=&\frac{(q^2-1)  \left(191 q^4+23 q^2+2\right) \left[(x^1) ^2-5 (x^2) ^2\right]}{5040 \pi ^2 q^6 r^8}, 
\\ \nonumber
\langle T_{22}\sigma_q \rangle &=&\frac{(q^2-1)  \left(191 q^4+23 q^2+2\right) \left[-5 (x^1)^2+(x^2) ^2\right]}{5040 \pi ^2 q^6 r^8}, 
\\ \nonumber
\langle T_{12} \sigma_q \rangle
&=&\frac{\left(-191 q^6+168 q^4+21 q^2+2\right) x^1x^2}{840 \pi ^2 q^6 r^8}.
\end{eqnarray}
From the above equations and also the component $T_i^i$ in (\ref{tstress2}),  we read out 
\begin{equation}
h_q = \frac{(q^2-1) ( 191 q^4 +23 q^2 +2)}{ 5040\pi^2 q^5}.
\end{equation}
This value coincides with the result obtained using the partition function on 
$S^1\times AdS_5$ in (\ref{2formhq}). 
Let us also obtain the first and second derivatives of $h_q$ 
\begin{equation}
h_q^{\prime}|_{q=1} = \frac{3}{35 \pi^2}, \qquad h^{\prime\prime}_q|_{q=1} = - \frac{9}{70\pi^2}.
\end{equation}

\subsection{Conformal forms in  $D$ dimensions} \label{d2form}

Our analysis of the $2$ form in $D=6$ shows that the gauge invariant partition function 
on $S^1\times AdS_5$ reduces to the partition function of a  Kaluza-Klein tower of 
co-exact $2$ forms on $AdS_5$. 
This  just depends only on the Plancherel measure of co-exact forms. 
In this section we use this observation and evaluate the entanglement entropies 
as well as twist operator weights for conformal forms in even $D$ dimensions. 

The R\'{e}nyi entropies are given by the 
expression 
\begin{equation}
S_q =\frac{1}{1-q} \int d\mu_{{\frac{d-1}{2}-\rm{form}} }(\lambda) \left[ q \log( 1- e^{-2\pi  \lambda} ) - 
 \log( 1- e^{-2\pi q \lambda} ) \right], 
\end{equation}
where the Plancherel measure of the $(d-1)/2 $-form on $AdS_d$ is given by 
\begin{eqnarray}
d\mu_{\frac{d-1}{2}-{\rm form }} (\lambda) &=& \frac{\hat g(d) {\rm Vol} (AdS_d )}
{ 2^{d-1} \pi^{\frac{d}{2} } \Gamma( \frac{d}{2} ) } \prod_{ j =1}^{\frac{d-1}{2}  } ( \lambda^2 + j^2) , 
\\ \nonumber
\hat g(d) &=& \frac{ ( d-1)!}{ (  \Gamma( \frac{d+1}{2} ) )^2 }.
\end{eqnarray}
The weight of the twist operator is given by 
\begin{eqnarray} \label{hqconform}
h_q &=& \frac{q \hat g(d) }{ 2^{d-1} \pi^{\frac{d}{2} - 1}  d \Gamma( \frac{d}{2} ) } 
\int_0^\infty d\lambda  \lambda \prod_{ j =1}^{\frac{d-1}{2}  } ( \lambda^2 + j^2) ( \coth \pi \lambda -\coth\pi q \lambda) . \nonumber \\
\end{eqnarray}
Appendix \ref{appen3}
 lists the  R\'{e}nyi entropies and $h_q$ for all even dimensions up to $4\leq D\leq14$. 
In table  \ref{table2} we list the entanglement entropies, the first and second derivatives of 
$h_q$ for these dimensions.

\begin{table}[ht]
\centering { \footnotesize{
\begin{tabular}{c|c|c|c}
\hline
$D $ & Coefficient of $\log(\epsilon)$ & $ h_q'|_{q=1}$  & $ h_q^{\prime\prime}|_{q=1}$  \\
\hline 
& &  \\
$4$ & $\frac{ 16}{45} $ & $\frac{1}{15\pi} $ & $- \frac{4}{45\pi} $
   \\   
&  &  & \\
$6$ &  $ -\frac{229}{630} $ & $ \frac{3}{35\pi^2}$  &  $-\frac{9}{70\pi^2}$ \\
& &  &  \\
$8$ & $\frac{1042}{2835} $ &  $\frac{4}{21\pi^3} $  & $-\frac{32}{105\pi^3} $\\
&  &  & \\
$10 $& $-\frac{276929}{748440}$ & 
$\frac{20}{33\pi^4} $  & $-\frac{100}{99\pi^4} $  \\
& &  &  \\
$12$& $\frac{45201643}{121621500}$
& $ \frac{360}{143\pi^5} $ & $-\frac{4320}{1001\pi^5}$ \\
& & & \\ 
$14 $& 
$-\frac{108829363}{291891600}$
&  $\frac{168}{13\pi^6} $  & $-\frac{294}{13\pi^6} $\\
&  & & \\
\hline
\end{tabular}
\caption{Entanglement entropies  conformal $\frac{D}{2}-1$-forms. The derivatives of 
$h_q$ agree with \ref{resultderhq} obtained using conformal invariance. }
\label{table2}
}}
\end{table}

\subsubsection*{$h_q$ and the 3 point function of the stress tensor}

Using conformal invariance it was shown in \cite{Hung:2014npa}, that the first and second derivatives of 
the conformal dimension of the twist operator $h_q$  with respect to $q$  at $q=1$are related to the 
parameters determining the 2 and 3 point functions of the stress tensor. 
This relation is known to be violated for scalars due to total derivative terms in the stress tensor
\cite{Lee:2014zaa,Herzog:2016bhv}.   In section \ref{maxfield} we have seen that these relations hold for the Maxwell field. 
Therefore it is  interesting to verify if such relations exist for the 
$2$ form. 

Before we proceed let us present the general relations in arbitrary $D$ dimensions.
The first and second derivatives of $h_q$ are expected to obey the relations
\begin{eqnarray}\label{hqabc}
& & h_q^\prime|_{q=1}=\frac{\pi ^{\frac{D+3}{2}} C_T}{2^{D-3} D \left(D^2-1\right) \Gamma \left(\frac{D-1}{2}\right)} , \quad
C_T =\frac{\left(8  \pi ^{D/2}\right) (a (D-2) (D+3)-2 b-c (D+1))}{D (D+2) \Gamma \left(\frac{D}{2}\right)},
\nonumber \\
& &  \nonumber
h_q^{\prime\prime} |_{q=1} =-\frac{16 \pi ^{D+1} }{D^2 \Gamma (D+3)}
 \left[2 a \left(3 D^2-3 D-4\right) (D-2)-2 b D (D-1)-c (3 D-4) (D+1)\right],
 \\
\end{eqnarray}
where $a, b, c$ are the parameters determining the $3$ point functions of the stress tensor
\cite{Osborn:1993cr}.
For the theory of free $\frac{D}{2} -1$-form in $D$ dimensions, the values of $a, b, c$ were evaluated 
in \cite{Buchel:2009sk}. These are given by 
\begin{eqnarray}
& & a=-\frac{D^3 \Gamma (D-1) \Gamma \left(\frac{D}{2}\right)}{64 (D-3) \pi ^{\frac{3 D}{2}}}, 
\qquad
b=-\frac{(D-4) D^3 \pi ^{-\frac{3 D}{2}} \Gamma (D-1) \Gamma \left(\frac{D}{2}\right)}{64 (D-3)},
\nonumber \\
& & c=-\frac{(D-2) D^3 \pi ^{-\frac{3 D}{2}} \Gamma (D-1) \Gamma \left(\frac{D}{2}\right)}{32 (D-3)}.
\end{eqnarray}
Now substituting these in the expressions for the derivatives of $h_q$ in (\ref{hqabc}) we obtain 
the following predictions for the behaviour of conformal dimensions of 
 twist operators of the  $\frac{D}{2}-1$ forms 
\begin{eqnarray}\label{resultderhq}
h_q^{\prime} |_{q=1} =
\frac{  D \Gamma( D-1)}{ 2^{ D+2}  \pi^{ \frac{D-3}{2} }  \Gamma ( \frac{ D+3}{2}) } , \qquad
 h_q^{\prime\prime}|_{q =1}  = -\frac{ D^3 \Gamma ( D-1) \Gamma( \frac{D}{2} ) }{
2 \pi^{ \frac{D}{2} - 1} \Gamma( D+ 3) }.
\end{eqnarray}
In the 3rd and 4th column of table \ref{table2} we have evaluated these derivatives
directly from the evaluation of $h_q$ using the expression in (\ref{hqconform}) which
relies on the partition function of conformal forms on hyperbolic cylinders. 
These values precisely agree with that obtained using conformal invariance in (\ref{resultderhq}).

 \section{Symmetric tensors}\label{tenfield}

In this section we evaluate the partition function of symmetric tensors of rank-$s$ on 
hyperbolic cylinders.
From the analysis of the Maxwell field and 2-form in $D=6$ in the previous sections, 
it wold be natural  to consider the 
partition function of a Kaluza-Klein tower of massive transverse symmetric traceless  spin-2
fields on $AdS_3$. 
The mass of the zero mode on $S^1$  should saturate the  Breitenlohner-Freedman bound.
However since the action of the  spin-2 field is not conformal invariant, it is certainly not expected 
such a partition function will indeed capture the entanglement entropy of the spin-2 field as 
evaluated by \cite{Benedetti:2019uej}. 
In this section we evaluate this partition function and provide arguments to support that 
this is indeed the natural partition  function to consider and argue why the result of this 
partition function should coincide with the result for the 
entanglement entropy for the spin-2 field  obtained by  \cite{Benedetti:2019uej}. 
We generalise this discussion for the spin-$s$ field in $D=4$. 
Using the recent results for the edge modes for the partition function of spin-$s$ fields on spheres
found by \cite{Anninos:2020hfj}, 
we show that the result for the entanglement entropy evaluated from the 
partition function of the tower of  Kaluza-Klein modes 
of spin-$s$ fields on $S^1\times AdS_3$  together with the contribution of the edge modes 
precisely agrees with that evaluated using the 
 partition function of the spin-$s$ field on $S^4$. 
 This agreement  
 provides a consistency check for using the Kaluza-Klein tower of massive traceless-transverse 
 spin-$s$ field to evaluate entanglement entropy across  a spherical entangling surface. 
 
 Finally for completeness  and future reference we generalise these calculations to 
 $D= 6, 8, 10, 12$.

\subsection{Gravitons in $D=4$}
 \label{secgraviton}

The linearised action of gravitons in flat space  is given by
 \begin{eqnarray}\label{spin2act}
S=  \int d^4x \left[ - \partial_\mu h^{\mu \nu} \partial_\alpha h^\alpha_{\; \nu} 
+ \frac{1}{2} \partial^\alpha h_{\mu\nu} \partial_\alpha h^{\mu\nu}
+ \partial_\mu h^{\mu\nu} \partial_\nu h^\alpha{\;\alpha}
- \frac{1}{2} \partial_\alpha h^\mu_{\;\; \mu} \partial^\alpha h^\nu_{\;\; \nu}
\right]
\end{eqnarray}
After fixing the gauge, and canonically quantising in radial co-ordinates \cite{Benedetti:2019uej}
 evaluated the 
entanglement entropy of a spherical entangling surface. 
The difficulty in using the  linearised action of gravitons  on $S^1\times AdS_3$ to evaluate 
the entanglement entropy are the following:
\begin{itemize}
\item
The action (\ref{spin2act}) is classically not conformal invariant so it is is not clear that
it is related to the theory of the spin-2 field on $S^1\times AdS_3$. 
\item 
$S^1\times AdS_3$  is not a solution to the  Einstein equation with a given cosmological 
constant. Therefore one would not obtain a quadratic action by considering 
spin-2 fluctuations about  the backgound $S^1\times AdS_3$. Indeed one has to 
set the fluctuations along $S^1$ to be zero by hand if  the cosmological  
constant  is chosen so that the spatial components  of the 
 Einstein equations are satisfied.   
\end{itemize}
Instead we  consider the action in (\ref{spin2act}) on $S^1\times R^3$,  fix gauge and 
obtain the partition function.  This is done in appendix \ref{appen4}
We are led to evaluate the following gauge invariant  partition function
\begin{eqnarray}
{\cal Z} = \int {\cal D} h_{ij}^\perp \exp\left[
-\frac{1}{2}  \int d^4 x  h^{\perp\;ij} ( - \partial_\tau^2 - \partial^k \partial_k ) h_{ij}^\perp
\right].
\end{eqnarray}
Here the derivative in the temporal direction picks out the Kaluza-Klein mass of $h_{ij}^\perp$ which 
is a transverse traceless spin-2 field  on $R^3$ satisfying 
\begin{equation}\label{ttt}
\partial^i h_{ij}^\perp =0, \qquad  h^i_{\; i} = 0.
\end{equation}
We now turn on the curvature in $R^3$, this  will introduce curvature couplings. 
The principle we use to introduce the curvature couplings is the following.
In $3$ dimensions we can write the flat space Laplacian using the identity
\begin{eqnarray}
h_{\;k}^{\perp\;n} \epsilon^{kl}_{\;\;m} \epsilon^{mij} \partial_l \partial_i h_{j n}^\perp
&=& h_{\;k}^{\perp\; n } (\delta^{ki} \delta^{lj} - \delta^{kj} \delta^{li}) \partial_l \partial_i h_{j n}^\perp,
\\ \nonumber
&=& - h^{kn\;\perp} ( \partial^i \partial_i) h_{kn}^\perp.
\end{eqnarray}
Note that in arriving at the identity, the traceless condition in (\ref{ttt}) has been used. 
Therefore on $R^3$ we can re-write the  flat space Laplacian  by the square of  the operator
$\epsilon^{mij} \partial_i $ 
Thus the partition function is rewritten as
\begin{eqnarray}
{\cal Z} = \int {\cal D} h_{ij}^\perp \exp\left[
-\frac{1}{2}  \int d^4 x  \left(  -h^{\perp\;ij}  \partial_\tau^2 h_{ij}^\perp  + 
h_{\;k}^{\perp\;n} \epsilon^{kl}_{\;\;m} \epsilon^{mij} \partial_l \partial_i h_{j n}^\perp 
\right) 
\right].
\end{eqnarray}
To introduce curvature couplings we
we now  promote the ordinary derivatives to covariant derivatives on $AdS_3$,  this results in the 
\begin{eqnarray} \label{covman}
&&h_{\;k}^{\perp\; n} \epsilon^{kl}_{\;\;m} \epsilon^{mij} \nabla_l \nabla_i h_{j n}^\perp
=  h_{\; k}^{\perp\;n} (g^{ki} g^{lj} - g^{kj} g^{li}) \nabla_l \nabla_i h_{j n}^\perp, \\ \nonumber
&=& h_{\; k}^{\perp\; n} g^{ki} g^{lj}\nabla_l \nabla_i h_{j n}^\perp  - h^{ \perp\; kn} \nabla^2 h_{kn}^\perp,  \\ \nonumber
&=& h_{\;k}^{\perp\; n } g^{ki} g^{lj}( \nabla_i \nabla_l  h_{jn} 
+ R_{jmli} h^{m \perp}_{\;\; n}  + R_{nmli} h_{j}^{\;\;m})   - h^{ \perp\; kn} \nabla^2 h_{kn}^\perp , \\ \nonumber
&=& - 2 h^{\perp\; kn }h_{kn}^\perp -  h^{\perp\; kn } h_{kn}^\perp  - h^{ \perp\; kn} \nabla^2 h_{kn}^\perp.
\end{eqnarray}
Here the $\epsilon$ tensor in curved space  along the $3$ directions  is defined by 
\begin{equation}
\epsilon^{ijk}  = \frac{\tilde \epsilon^{ijk}}{\sqrt{g}} , \qquad \tilde\epsilon^{123} = 1
\end{equation} 
To obtain the last line in (\ref{covman})  we have used $\nabla^i h_{ij}^\perp =0$, 
the traceless condition
 $h^{i \perp}_i = 0$ and the Riemann curvature components  of $AdS_3$ given in (\ref{curvads3}). 
 It is important to 
 note that this principle naturally determines a  definite value of the  curvature induced mass
 \footnote{We will subsequently see that the introducing the 
 curvature couplings using this principle results in a scale invariant partition function.
 The intuition for introducing curvature couplings arises from the fact that higher spin equations
 of motion in  $3$ dimensions can be written as square of the operator involving 
  $\epsilon^{mij} \nabla_i$ \cite{Tyutin:1997yn,Datta:2011za}}. 
 Therefore the partition  function of the spin-$2$ field on $S_1\times AdS_3$ is given by 
  \begin{eqnarray} \label{parttgrav}
 {\cal Z} &=&  \int {\cal D} h_{ij}^\perp \exp\left[
-\frac{1}{2}  \int \sqrt{g} d^4 x   h^{\perp\; ij} ( - \partial_\tau^2 - \nabla^2 - 3) h_{ij}^\perp
\right] \\ \nonumber
 &=& \left[  \frac{1}{ {\rm det} (  -\partial_\tau^2 - \Delta_{(2)} - 3) }  \right]^{\frac{1}{2} }.
 \end{eqnarray}
 The action consists of a  Kaluza-Klein 
 tower of traceless transverse symmetric  spin-2 fields. 
 From (\ref{bfbound}) we see that the Kaluza-Klein zero mode satisfies the 
Breitenlohner-Freedman bound for spin-2 field in $d=3$.
 In fact as will be subsequently seen, performing the same analysis with  the traceless
 transverse spin-$s$ field, the mass for the Kaluza-Klein zero modes is given by 
 $m_s^2 = -(s+1)$ which satisfies the corresponding Breitenlohner-Freedman bound.

 We can now proceed as before and evaluate the free energy of this theory. 
 The eigenvalues of the spin-2 Laplacian on transverse, traceless spin-2 field
 on $AdS_3$  \cite{Camporesi:1994ga}
 \begin{equation}
 \Delta_{(2)} \psi_{i j}^{\{\lambda, u\}} = - ( \lambda^2 + 3) \psi_{ij}^{\{\lambda, u \}}.
 \end{equation}
 Therefore the free energy for the Kaluza-Klein tower of spin-2 Kaluza-Klein modes is given by 
 \begin{equation}
 {\cal F}_q = \frac{1}{2}\sum_{n} \int d\mu^{(2)} \log( \lambda^2  + \frac{n^2}{q^2} ) .
 \end{equation}
 The $m_s^2 =-3$ cancels the shift from the eigenvalue of the spin-2 Laplacian ensuring that 
 the Kaluza-Klein zero mode has a gapless spectrum. 
 The Plancherel  measure is given by  \cite{Camporesi:1994ga,Gopakumar:2011qs} 
 \begin{equation}\label{plans2}
 d\mu^{(2)} = \frac{{ \rm Vol }( AdS^3)}{\pi^2} (\lambda^2 + 4) d\lambda.
 \end{equation}
Here  again we have used the spin dependent factor $g(s)$ determined by 
 \cite{Gopakumar:2011qs}. 
 Substituting the Plancherel measure in the expression for the entanglement entropy 
 given in (\ref{defsq}), 
 we obtain 
 \begin{eqnarray}\label{gravrenyi}
 S_q &=& \frac{1}{ 1-q} \int d\mu^{(2)}[  q \log ( 1 - e^{ -2\pi \lambda}) - 
 \log( 1- e^{2\pi q\lambda}) ], \\ \nonumber
 &=& \frac{ ( q+ 1) ( 121 q^2 + 1) }{ 180 q^3} \log( \epsilon) .
 \end{eqnarray}
 We can now take the $q\rightarrow 1$ limit to obtain the entanglement 
 entropy, which results in 
 \begin{equation}
 S_1 = \frac{61}{45} \log(\epsilon).
 \end{equation}
 Note that this precisely coincides with the 
 result obtained by \cite{Benedetti:2019uej}. The method relied on  
 canonical quantization of the graviton
in spherical coordinates. 
In \cite{Dowker:2019zva} it was noted that the above result can also be obtained by considering the thermodynamics
in de-Sitter space which is conformally related to that of hyperbolic cylinder. 
This result   obtains the entanglement entropy by integrating the 
energy density in de-Sitter space.  However this depended on a reasonable 
definition of energy density which did not have a first principle derivation \cite{Dowker:1983nt}. 

We have shown the result follows from evaluating the partition function
 of  a  Kaluza-Klein tower of traceless transverse spin-2 fields 
$S^1 \times AdS_3$. 
The constant mode of the graviton on $S^1$ has the mass which saturates the 
Breitenlohner-Freedman  bound. 
The mass in  (\ref{parttgrav})  was introduced  using a definite principle for coupling 
the traceless transverse spin-2 fields to  $AdS_3$. At present we do not have a  first principle 
derivation 
that this theory is conformally related to that of the graviton in flat space. 
However,  it is easy to see  should the mass be any other mass other than 
$m_s^2|_{s=2} = - 3$, the partition function would not be just a function of the ratio of the 
length scales of $S^1$ and $AdS_3$.  To see this we re-instate the radius of $AdS_3$ and 
also introduce  an additional mass  for the spin-2 field. 
Thus the starting point is  given by 
 \begin{eqnarray}
 {\cal Z} &=&  \int {\cal D} h_{ij}^\perp \exp\left[
-\frac{1}{2}  \int \sqrt{g} d^4 x   h^{ij \; \perp} ( - \partial_\tau^2 - \nabla^2 - 3 +m^2 ) h_{ij}^\perp
\right] , \\ \nonumber
 &=& \left[  \frac{1}{ {\rm det} (  -\partial_\tau^2 - \Delta_{(2)} - 3 + m^2 ) }  \right]^{\frac{1}{2} }.
 \end{eqnarray}
 Evaluating the partition function using the eigen functions on $AdS_3$,  we obtain
\begin{eqnarray}
{\cal F}_q &=& \frac{1}{2} \sum_n \int d\mu^{(2)} \log( \frac{ \lambda^2}{R^2} +  m^2 + \frac{n^2}{q}) , 
\\ \nonumber
&=& 
\int d \mu^{(2)} \Big(\log(1-e^{-2\pi q\sqrt{\frac{\lambda^2}{R^2}+m^2}})+\pi q\sqrt{\frac{\lambda^2}{R^2}+m^2}\Big). 
\end{eqnarray}
We can measure the additional mass $m^2$ in terms of units of the radius of $AdS_3$ by introducing
$ \hat  m^2 = R^2 m^2$. 
Then the partition function becomes 
  \begin{align}\label{massivf}
                 \mathcal{F}_q&=\int d \mu^{(2)}
                 \Big(\log(1-e^{-2\pi \frac{q}{R}\sqrt{\lambda^2+\hat{m}^2}})+\pi \frac{q}{R}\sqrt{\lambda^2+\hat{m}^2}\Big) .
       \end{align}
Note the $\mu^{(2)}$ is given in (\ref{plans2}).  From (\ref{massivf}) we see that the free energy 
of the theory is a function ${\cal F}_q (  q/R, \hat m) $. 
The free energy of  a scale invariant theory on  $S^1\times AdS_3$ should be 
a function only of the ratio of the the  radius of $S^1$ to  $AdS_3$. 
From  (\ref{massivf}), we see that this  property holds only if and only if 
$\hat m =0$.  Thus we have shown
that it is only when the mass saturates the Breitenlohner-Freedman bound, the theory is 
scale invariant.

Finally we can proceed to evaluate the dimension of the twist operator 
localized on the entangling surface using the expression in (\ref{condimtwist}). 
This results in 
\begin{eqnarray}\label{gravhq}
h_q&=& \frac{q}{3\pi}\int_0^\infty  d\lambda \, 
\lambda (\lambda^2 +4) [ \coth (\pi \lambda) - \coth( \pi q \lambda) ], \\ \nonumber
&=& \frac{(q^2-1)( 41q^2+1) }{ 360 \pi q^3} .
\end{eqnarray}
Examining the expressions for the R\'{e}nyi entropy and $h_q$ in (\ref{gravrenyi}) and 
(\ref{gravhq}) and their counterparts for the spin-0 and spin-1 discussed earlier, 
we observe that  R\'{e}nyi entropy and the conformal dimensions 
of the twist operator are certain  integral transforms of the Plancherel measure of 
the corresponding  field.

 \subsection{Higher spin fields in $D=4$ }\label{sechighspin}
 
 Just as in the case of the graviton we can begin with the action of higher spin fields 
 in flat space, fix gauge and arrive at the following gauge invariant action partition function 
 for traceless transverse
 spin-$s$ tensors on $S^1\times R^3$. 
 \begin{eqnarray}\label{hsact}
{\cal Z} = \int {\cal D} h_{ij}^\perp \exp\left[
-\frac{1}{2}  \int d^4 x  h^{\perp\; i_1 i_2, \cdots i_s} ( - \partial_\tau^2 - \partial^k \partial_k ) 
h_{i_1i_2, \cdots i_s}^\perp
\right].
\end{eqnarray}
where $h_{i_1, \cdots i_s}^\perp$ are totally symmetric tensors satisfying 
\begin{equation}\label{ttt2}
\partial^i h_{i , i_2, \cdots i_s}^\perp =0, \qquad h^{\perp\;\;i}_{i i_2\cdots i_s} =0.
\end{equation}
We now introduce curvature couplings through the same principle. 
We can re-write the Laplacian  in (\ref{hsact})  in terms of the $\epsilon $ tensor as 
\begin{eqnarray}
h_{k}^{ \perp\; \; n_2, \cdots n_s} \epsilon^{kl}_{\;\;m} \epsilon^{mij} \partial_l \partial_i h_{j n_2, \cdots n_s }^\perp
&=& h_{k}^{\perp\;\; n_2, \cdots n_s } 
(\delta^{ki} \delta^{lj} - \delta^{kj} \delta^{li}) \partial_l \partial_i h_{j n_2, \cdots n_s }^\perp
\\ \nonumber
&=& - h^{\perp\;\; j_1 j_2, \cdots j_s } ( \partial^i \partial_i) h_{j_1 j_2, \cdots j_s }
\end{eqnarray}
In  arriving at the identity, the traceless condition in (\ref{ttt2}) has been used. 
As before we 
 promote the ordinary derivatives to covariant derivatives on $AdS_3$,  this results in the 
\begin{eqnarray}
&& h_{k}^{\perp\;\; n_2, \cdots n_s} \epsilon^{kl}_{\;\;m} \epsilon^{mij} 
\nabla_l \nabla_i h_{j n_2\cdots n_s}^\perp, 
=  h_{k}^{\perp\;\;n_2 \cdots n_s} 
(g^{ki} g^{lj} - g^{kj} g^{li}) \nabla_l \nabla_i h_{j n_2 \cdots n_s }^\perp,  \\ \nonumber
&=& h_{k}^{\; n\;\perp} g^{ki} g^{lj}\nabla_l \nabla_i h_{j n}^\perp  - h^{kn\; \perp} \nabla^2 h_{kn}^\perp, 
 \\ \nonumber
&=& h_{k}^{\perp\;\; n_2 \cdots n_s } g^{ki} g^{lj}( \nabla_i \nabla_l  h_{jn_2\cdots n_s} ^\perp
+ R_{jmli} h^{m \perp}_{\;\; n_2\cdots n_s}  + 
R_{n_2 mli} h_{j\;\;\; n_3\cdots n_s }^{\perp\;\;m} 
+ R_{n_3 mli} h_{jn_2\;\;\;  n_4\cdots n_s }^{\perp\;\;m}  + \cdots 
) \\ \nonumber
& & \qquad
-  h^{\perp\;\;i_1\cdots i_s} \nabla^2 h_{i_1  \cdots i_s}^\perp, \\ \nonumber
&=& - 2 h^{\perp \; i_1 \cdots i_s }h_{i_1 \cdots i_s }^\perp -  (s-1) h^{\perp\; i_1 \cdots i_s} 
h_{i_1 \cdots i_s }^\perp -
 h^{\perp\;\;i_1\cdots i_s} \nabla^2 h_{i_1 \cdots i_s}^\perp. 
\end{eqnarray}
To obtain the last line we have used $\nabla^j h_{ji_2\cdots i_s }^\perp =0$ and the traceless condition
 $h^{\perp \; j }_{j\i_2\cdots i _s}= 0$ \footnote{As  discussed for the spin-2 field, 
 introducing curvature couplings using this principle results in a scale invariant partition function
 also for the spin-$s$ fields. 
 In fact the action $S = \int d^3 \sqrt{g} h^{\perp\;i_1 \cdots i_s} \epsilon^{\;\; jk}_{i_1} \nabla_j h^\perp_{ki_2\cdots i_s}$ is invariant under the Weyl transformation $g_{ij} \rightarrow e^{2\phi} g_{ij}$
 and $h^{\perp}_{i_1 \cdots i_s } \rightarrow e^{ (s-1) \phi} h^{\perp}_{i_1 \cdots i_s }$. 
 The Chern-Simons action is a very special case of this action which is topological. 
 This perhaps might be the reason that the curvature couplings induced using this principle 
 result in a scale invariant partition function. 
 }
 
 The partition function for these tower of Kaluza-Klein traceless transverse spin-$s$ particles are given 
 by 
   \begin{eqnarray}
 {\cal Z} &=&  \int {\cal D} h_{ij}^\perp \exp\left[
-\frac{1}{2}  \int \sqrt{g} d^4 x   h^{ \perp\;\; i_1 \cdots i_s } ( - \partial_\tau^2 - \nabla^2 - (s+1) ) h_{ij}^\perp
\right] \\ \nonumber
 &=& \left[  \frac{1}{ {\rm det} (  -\partial_\tau^2 - \Delta_{(2)} - (s+1)) }  \right]^{\frac{1}{2} }.
 \end{eqnarray}
 Again from (\ref{bfbound}) we see that mass of  the Kaluza-Klein zero mode on $S^1$ 
 given by $m_s^2 = - ( s+1)$ satisfies the 
 Brietenlohner-Freedman bound of the spin-$s$ field in $d=3$. 
 We can evaluate the free energy 
using eigenfunction of the spin-$s$ Laplacian  whose eigenvalues are given in (\ref{eigenspins}).
 This results in 
 \begin{eqnarray}\label{parths}
  {\cal F}_q &= & \frac{1}{2}\sum_{n} \int d\mu^{(s)} \log( \lambda^2  + \frac{n^2}{q^2} ) , 
  \\ \nonumber
 d\mu^{(s)} &=& \frac{{\rm Vol} (AdS_3)}{\pi^2} ( \lambda^2 + s^2) d\lambda.
 \end{eqnarray}
 Here we have taking the Plancherel measure from \cite{Camporesi:1994ga, Gopakumar:2011qs}.
 Evaluating the R\'{e}nyi entropy using (\ref{defsq}) we obtain
 \begin{eqnarray}\label{renyihs}
 S_q &=& \frac{1}{1-q} \int d\mu^{(s)} \left[ q \log ( 1 - e^{ -2\pi \lambda}) - 
 \log( 1- e^{2\pi q\lambda}) \right], \\ \nonumber
&=& \frac{ ( q+ 1) [ ( 30 s^2 + 1)q^2 + 1] }{180 q^3 } \log \epsilon.
 \end{eqnarray}
 The entanglement entropy is  given by 
 \begin{equation}\label{spinsee}
 S_1  = \frac{ 15 s^2 + 1}{ 45} \log \epsilon.
 \end{equation}
 This result agrees with the conjecture of \cite{Benedetti:2019uej} as well as the result evaluated by 
 \cite{Dowker:2019zva} 
 using thermodynamic of de-Sitter space. 
 In addition it shows that the entanglement entropy is an integral transform of the Plancherel 
 measure. 
 The conformal dimension of the twist operator localised on the entangling surface is 
 given by 
 \begin{eqnarray}
 h_q &=& \frac{q}{3\pi} \int_0^\infty  d\lambda \, 
\lambda (\lambda^2 +s^2) [ \coth (\pi \lambda) - \coth( \pi q \lambda) ], \\ \nonumber
&=& \frac{( q^2 -1) ( q^2 ( 10 s^2 +1)  +1) }{ 360 \pi q^3}.
 \end{eqnarray}
 
 \subsection{Consistency check}
 
Let us  perform another consistency check on the  result of the entanglement entropy derived 
form using the theory of  the Kaluza-Klein tower of spin-$s$ particles on $S^1\times AdS_3$. 
We know that for conformal invariant theories, 
 the result from evaluating the 
the entanglement entropy  of the theory on  branched spheres  should coincide with that 
of hyperbolic cylinders modulo the edges modes. 
We have discussed earlier for the Maxwell theory, 
 the hyperbolic cylinders misses out the contribution of the edge modes, this same is true 
 for the conformal $D/2-1$-form theory in $D$ dimensions \cite{Dowker:2017flz}. 
 Let us check if this fact is true for the theory of  higher spin fields in $D=4$ dimensions. 
The coefficient of the 
 logarithmic divergence of the partition function for spin-$s$ fields on the $q$ branched sphere in 
$D=4$ has been 
evaluated in equation (E.4) of \cite{Beccaria:2017lcz}. 
This is given by \footnote{We have used the fact that  cutoff $\Lambda_{UV}$ in \cite{Beccaria:2017lcz}
is in momentum space and therefore related to a position space cut off $\epsilon \sim \frac{1}{\Lambda_{UV} }$. We thank Arkady  Tseytlin for bringing this reference to our attention 
after the first version of our manuscript  appeared on the arXiv. }
\begin{eqnarray}
&& F[q]=  -\log Z= \zeta(q,s), \\ \nonumber
     &=& -\frac{1}{180 q^3}+\frac{\left(1-s^2\right) s^2}{6 q^2}-\frac{\left(3-2 s^2\right) s^2}{6 q}-\frac{1}{60} q \left(30 s^4-20 s^2+1\right)+\frac{1}{6} \left(1-3 s^2\right) s^2. 
\end{eqnarray}
Using this we evaluate the logarithmic coefficient of the entanglement entropy as 
given by the partition function on the branched sphere. 
 \begin{eqnarray}
      S_q ( S^4_q) &=&\frac{ q F[1]-F[q] }{1-q}, \\ \nonumber
      &=&\frac{q \left(q^2+30 (q (2 q-1)+1) s^4+30 (q (q+2)-1) s^2+q+1\right)+1}{180 q^3}. \\ \nonumber
      S_1 ( S^4) &=&  \lim_{q\rightarrow 1} S_q( S^4_q) ,  \nonumber \\
       &=& \frac{1}{45} \left(15 \left(s^4+s^2\right)+1\right). 
  \end{eqnarray}
  Let us now evaluate the difference between the entanglement entropies evaluated by considering 
  the branched sphere and the theory of the Kaluza-Klein tower of   transverse spin-$s$ particles on $S^1\times AdS_3$ given in (\ref{spinsee}). 
  The logarithmic coefficient of the difference is given by 
  \begin{equation}\label{difference}
  S_1(S^4) - S_1(AdS_3\times S^1)=\frac{s^4}{3} . 
  \end{equation}
  This should coincide with the partition function of the edge modes. 
  Indeed, we see that for the Maxwell case,  $s=1$ it agrees with the 
  edge mode partition function found in \cite{Huang:2014pfa,Donnelly:2014fua}
  \footnote{See equation (46) of \cite{Huang:2014pfa}.}. 
  Recently the contribution of the edge modes for arbitrary spin in $D=4$ was evaluated in 
  \cite{Anninos:2020hfj}. 
  From this work we see that  logarithmic contribution to the partition function from the edge modes 
  for spin $s$ fields in $D=4, d=3$ 
  is obtained from 
  \begin{eqnarray}
    \log \mathcal{Z}_{edge}&=&\int \frac{dt}{2t}\frac{1+q}{1-q} ( - [\hat \chi_{{\rm edge}, s}]_+) 
    , \qquad q = e^{-t}, \\ \nonumber
    [\hat \chi_{{\rm edge}, s}]_+ &=&  \frac{1}{1-q} \left[ 
    \frac{1}{3} s (s+1) (2 s+1) q^s-\frac{1}{3} (s-1) s (2 s-1) q^{s+1} \right]. 
   \end{eqnarray}
   These are read out from  equation (5.11)  and the second line of equation 
( 5.9)  of \cite{Anninos:2020hfj}. 
To obtain the logarithmic coefficient we need to perform the Taylor series expansion  of the 
integrand in $t$ and extract out the coefficient of $1/t$, this results in 
\begin{equation}\label{edge}
S_{\rm edge} = \frac{s^4}{3}
\end{equation}
The contribution of the edge modes in (\ref{edge}) precisely agrees with the difference in 
the entanglement entropies  on the sphere and that given on $S^1\times AdS_3$ in (\ref{difference}). 
This agreement provides a strong consistency check that the theory 
of the tower of Kaluza-Klein traceless transverse spin-$s$ fields
 at the Brietenlohner-Freedman bound which includes the case of the graviton 
is related to the theory of higher spin fields  in flat space, as well as the theory of spin-$s$ 
on the sphere. 

Finally we would like to point out that the agreement  of entanglement entropy 
evaluated from the 
theory of 
 higher spin fields on $S^1\times AdS_3$ given in (\ref{spinsee}) with that 
 evaluated by  \cite{Benedetti:2019uej}  should not be a surprise given the structure of the 
 Plancherel measure of the higher spin fields on $AdS_3$. 
From (\ref{parths}) and (\ref{renyihs}), 
 we see that spin dependence arises only from the Plancherel measure
and it is quadratic in spin, thus the logarithmic coefficient of the entanglement entropy is a function 
of the form
\begin{equation}
S_1( S^1\times AdS_3) = A+ B s^2.
\end{equation}
The mode counting argument of Benedetti and Casini in  \cite{Benedetti:2019uej} 
also show that the entanglement entropy is also of form 
\begin{equation}
S_1( {\rm BC})  = A' + B' s^2.
\end{equation}
  For spins $s=0,  s=1$, the theories are 
are conformally invariant and their  actions on conformally flat spaces are related. 
Now for $s=0$ and $s=1$, we know $S_1( S^1\times AdS_3)= S_1( {\rm BC})$, 
therefore  the agreement 
  for arbitrary spin-$s$ is not a surprise.

\subsection{Symmetric tensors in arbitrary $D$}\label{secarbD}

In this section  for completeness and future reference, 
we evaluate the free energy of the Kaluza-Klein tower of symmetric 
traceless rank-$s$ tensors on $AdS_d$, 
$D= d+1$ is even. 
The mass of these fields saturate the Breitenlohner-Freedman bound. 
We then use the expression (\ref{defsq}) to evaluate the  R\'{e}nyi entropies and entanglement
entropy. 
The result for all even  $D$ dimensions is given by 
\begin{equation}
S_q =  \frac{1}{1-q} \int d \mu^{(s)} 
[  q \log ( 1 - e^{ -2\pi \lambda}) - 
 \log( 1- e^{2\pi q\lambda}) ], \\ \nonumber
\end{equation}
where the Plancherel measure is given by  \cite{Camporesi:1994ga,Gopakumar:2011qs}
\begin{eqnarray} \label{gs}
 \mu^{(s)} (\lambda) d\lambda
 &=& \frac { {\rm Vol } ( AdS_d )g(s) }{ 2^{d-1} \pi^{\frac{d}{2} } \Gamma( \frac{d}{2} ) }
\left[ \lambda^2 +(s+\frac{d-3}{2})^2\right]
\prod_{j =0}^{ \frac{d-5}{2}} ( j^2 + \lambda^2) d\lambda,  \\ \nonumber
g(s) &=& \frac{(2s+d-3)\Gamma(s+d-3)}{\Gamma(d-2)s!}.
\end{eqnarray}
The spin dependent factor $g(s)$ was determined in \cite{Gopakumar:2011qs} using 
analytical continuation of the heat kernel on $S^d$. 
For $d=3$,  $g(0) =1$ and $g(s) =2$ for all $s\geq1$. As we have seen in the 
previous section that fact that this factor of $2$ is important for the result of the 
entanglement entropy of  gravitons  to precisely agree with that obtained in \cite{Benedetti:2019uej}. 
For $d>3$,   $g(s)$ depends non-trivially  on spin. 
The regularized volume of $AdS_d$  is given in  \cite{Hung:2011nu}. 
\begin{equation} \label{voladsd1}
{\rm Vol} ( AdS_d )  =  \frac{ 2 \pi^{\frac{d-1}{2} } ( -1)^{ \frac{d+1}{2} } }{ \Gamma( \frac{d+1}{2} ) }
\log( \epsilon) .
\end{equation}
Table  \ref{table1}. lists the coefficient of the logarithm for entanglement entropies
  of the spin-$s$ field for all even $D$ 
dimensions $4\leq D\leq 14$. The last column of this table takes the $s\rightarrow 0$ limit. 
The values agrees with the result for the entanglement entropy of a conformally coupled scalar across spheres \cite{Casini:2010kt}. 
\begin{table}[ht]
\centering { \footnotesize{
\begin{tabular}{c|l|c}
\hline
$D $ & Coefficient of $\log(\epsilon)$ & $s\rightarrow 0$ \\
\hline 
& &  \\
$4$ & \quad$\frac{ 1+ 15 s^2}{45} $ & $\frac{1}{90}\footnotemark$
   \\   
&  & \\
$6$ & \quad $- \frac{( s+1)^2 ( 7 s^2 + 14 s + 10)}{7560 s!}$ & $ -\frac{1}{756}$  \\
& &  \\
$8$ & \quad$ \frac{ ( s+1) ( s+2)^2 ( s+3) ( 5s^2 + 20 s + 23)}{1360800}  $ &  $\frac{23}{113400}$  \\
&  & \\
$10 $& \quad$- \frac{ ( s+3)( 759 s^2 + 4554 s + 7364 ) ( s+ 5)! }{75442752000 s!} $ &
$ -\frac{263}{7484400}$  \\
& &  \\
$12$& \quad $\frac{( s+4) ( 23933s^2+191464s + 401361) (s+7)! }{1235752277760000s! } $
& $\frac{133787}{20432412000} $\\
& & \\ 
$14 $& \quad 
$-\frac{(s+5) \left(133787 s^2+1337870 s+3454198\right) (s+9)!}{4893579019929600000 s!}$
&  $- \frac{157009}{122594472000} $\\
&  & \\
\hline
\end{tabular}
\caption{Entanglement entropies of spin-$s$ fields in $D$ dimensions. }
\label{table1}
}}
\end{table}
\footnotetext{ For $D=4$ we have taken the limit $s\rightarrow 0$ first and then taken
$D\rightarrow 4$. The order of limits matters for this case due to the nature of the function $g(s)$ in 
(\ref{gs}). 
}

The R\'{e}nyi entropies are listed in appendix \ref{appen3}.
Finally we can also evaluate evaluate the conformal dimensions of the twist operator 
localized on the entangling surface using (\ref{condimtwist}). 
This leads to the following integral
\begin{eqnarray}
h_q = \frac{q g(s) }{ 2^{d-1} \pi^{\frac{d}{2} - 1}  d \Gamma( \frac{d}{2} ) } 
\int_0^\infty d\lambda \lambda \left[ \lambda^2 +(s+\frac{d-3}{2})^2\right]
\prod_{j =0}^{ \frac{d-5}{2}} ( j^2 + \lambda^2) [ \coth (\pi \lambda) - \coth( \pi q \lambda) ].
\nonumber \\
\end{eqnarray}
These polynomials  in $q$ for various dimensions  are listed in the appendix  \ref{appen3}

\section{Conclusions}\label{conclude}

We have evaluated the partition function of the  $U(1)$  
  in $D=4$ and the 2-form theory 
in $D=6$
starting from a gauge invariant action on the hyperbolic cylinder. 
After fixing gauge  by adapting the functional methods of 
\cite{Fradkin:1983mq,David:2009xg,Gaberdiel:2010ar}, we demonstrate that
for these cases, the partition function reduces to a Kaluza-Klein tower of 
states with the corresponding spin in $AdS$ space. 
For the conformal $U(1)$ theory and the 2-form theory, the mass of the 
constant mode along $S^1$ satisfies the Brietenlohner-Freedman bound.
We have verified that the entanglement entropy as well conformal dimensions of the 
co-dimension 2 twist operator of the $U(1)$ and the 2-form theory satisfies known  consistency checks.

The spin-2 theory in $D=4$ is not conformally invariant, we begin with the 
gauge invariant partition function of gravitons on $S^1\times R^3$ and 
 using a principle to introduce curvature couplings, we obtain the
partition function of the Kaluza-Klein tower of traceless  transverse spin-2 fields. 
Again the mass of the constant mode along $S^1$ saturates the Brietenlohner-Freedman bound. 
This partition function is show to be scale invariant. 
The  logarithmic coefficient of the entanglement entropy evaluated from this 
for the spin-2 theory agrees with 
that obtained by \cite{Benedetti:2019uej}.   
We repeat this analysis for higher spin fields in $D=4$ and show that the result coincides with the 
conjecture in \cite{Benedetti:2019uej}. 
Since the theory of gravitons in $D=4$ is not conformal 
invariant , as a  consistency  check on the application of the hyperbolic cylinder approach to 
the case of gravitons we demonstrate that the edge modes for arbitrary spin fields obtained from this 
approach precisely coincides with that recently obtained in \cite{Anninos:2020hfj}. 

In all the three cases, 
both the entanglement entropy and the weight of the twist operator are obtained by 
simple integral transforms of the Plancherel measure of the particle with the corresponding
spin in $AdS$ space. 
This resulted from the fact  that the partition functions  consisted of a tower of Kaluza-Klein 
fields  on the hyperbolic cylinder 
with  corresponding spin whose constant modes on $S^1$ satisfied the 
Brietenlohner-Freedman bound. 
It will be interesting to generalise this discussion to higher spin fermions and observe 
if the same conclusions hold.   One can adapt the functional methods 
applied for the gravitino in 
\cite{David:2009xg}.

The logarithmic coefficient of the entanglement entropy for the $U(1)$ theory and the $2$-form theory
obtained by evaluating the partition function on the hyperbolic cylinder do not agree with the 
expected  trace anomaly coefficient.  
The shift to restore this coefficient to the anomaly is attributed to the edge modes or  the extended
Hilbert space. 
Recently, in \cite{Casini:2019nmu} it was shown that indeed, for the free Maxwell 
field without  the presence of charged states, it is possible for this coefficient to be different from that 
of the trace anomaly.  It was further shown that coupling with heavy charges changes this
coefficient to that of the trace anomaly. 
It will be interesting to repeat this exercise for graviton or the 2-form theory to understand this 
phenomenon further.

\appendix

\section{Change in measures}\label{appen1}

In this appendix we evaluate the change in the functional integration measure 
for the various change of variables used in performing the path integrals in the main text.
We follow the methods developed in \cite{Fradkin:1983mq}. 

\subsection*{Maxwell field in $D=4$}
From the canonical measure  ${\cal D} A_\mu$ we have the result 
  \begin{align}\label{measureMax}
  \int \mathcal{D} A_{\mu} e^{-\int d^4 x \sqrt{g}  A_{\mu} A^{\mu}}=1.
  \end{align}
  Let us change variables to 
  \begin{equation}\label{trans1}
  \{A_\mu \} \rightarrow  \{ A_\tau, A_i^\perp, \phi\} , 
  \end{equation}
where $i$ refers to 
  the $AdS_3$ directions and $\nabla^i A_i^\perp =0$. 
  We can write the term in the exponent  as 
  \begin{equation}
  \int d^4 x \sqrt{g}  A_{\mu} A^{\mu} = \int d^4 x \sqrt{g} ( A_\tau^2 + A_i^\perp A^{\perp \, i } + 
  \phi ( - \Delta_{(0)} ) \phi ) .
  \end{equation}
  We have used integration by parts and the transversal property of $A_i^\perp$ to arrive at the 
  above equation.   $\Delta_{(0)}$ is the scalar Laplacian on $AdS_3$. 
  Let determinant of the Jacobian involved in the change of measure be $J$. 
  Then from (\ref{measureMax}) we obtain
  \begin{equation}
  \int {\cal D} A_\tau {\cal D} A_i^\perp {\cal D}\phi  J 
  e^{ - \int d^4 x \sqrt{g} ( A_\tau^2 + A_i^\perp A^{\perp \, i } + 
  \phi ( - \Delta_{(0)} ) \phi ) }  =1.
  \end{equation}
  Now performing all the functional integrals, we obtain
  \begin{equation}
  J [ {\rm det} (- \Delta_{(0) }) ]^{- \frac{1}{2} } = 1.
  \end{equation} 
  Therefore the change in measure for the transformation given in (\ref{trans1}), we obtain 
  \begin{equation}\label{max-changem}
   \mathcal{D} A_{\mu} =  {\cal D} A_\tau {\cal D} A_i^\perp {\cal D}\phi \times J, \qquad
  J = [ {\rm det} ( -\Delta_{(0) }) ]^{ \frac{1}{2} } .
  \end{equation}

                                   \subsection*{Graviton in $D=4$: $S^1\times R^3$}
                                   
                                   The  fields for the graviton we have are $ h_{\tau\tau}, h_{\tau i }$ and $h_{ij}$. 
                                  We start with the canonical definition of he measure given by the integral
\begin{align}\label{gravnorm}
\int {\cal D} h_{\tau\tau}
\mathcal{D} h_{\tau i } {\cal D} h_{ij} e^{-\int d^4 x(   h_{\tau\tau}^2 + 2 h_{\tau i} h^{\tau i } + h_{ij} h^{ij}) 
}=1.
\end{align}
The coordinates  ${i,j}$ refer to  $R^3$ directions. 
We decompose $h_{ij}$ into its traces and traceless components by 
\begin{equation}
 h_{ij}=\bar{h}_{ij}+\frac{1}{3}\delta_{ij}h.
 \end{equation}
  The traceless component $\bar{h}_{ij}$ 
 can be further decomposed  as 
 \begin{eqnarray}
 & & \bar{h}_{ij}=h^{\perp}_{~ij}+\partial_{i}\zeta^{\perp}_j+\partial_{j}\zeta^{\perp}_ {i}+\partial_{i}\partial_{j}\sigma-\frac{1}{3}\delta_{ij}\Box\sigma, \\ \nonumber
 & & \partial^i h^{\perp}_{ ij}=0, \qquad\qquad \partial^i \zeta^{\perp}_{ i}=0.
 \end{eqnarray}
 With these definitions we have 
 \begin{eqnarray}
& &  \int d^4 x  h_{ij} h^{ij} =  \int d^4 x  \left[  h_{ij}^\perp h^{\perp\, ij}  + 
 (\partial_{i}\zeta^{\perp}_{ j}   +\partial_{j}\zeta^{\perp}_{i})^2  \right. \\ \nonumber
 & & \qquad\qquad\qquad\qquad \left. 
 + (\partial_i\partial_{j}\sigma-\frac{1}{3}\delta_{ij}\Box\sigma)^2+\frac{1}{3}h^2\right] .
 \end{eqnarray}
 The cross terms cancel due to integration by parts due to the transverse gauge condition as well 
 as the traceless condition. 
 Substituting this, 
 we can further simplify the terms by integration by parts and obtain
 \begin{eqnarray}\label{expansgrav}
 \int d^4 x \sqrt{g} h_{ij} h^{ij} &=&  \int d^4 x \sqrt{g} \left[  h_{ij}^\perp h^{\perp\, ij} 
 + 2 \zeta_{i}^{~\perp} (-\Delta_{(1)} )  \zeta^{i\perp}
 \right. \\ \nonumber
 &&\left.  +\frac{2}{3}\sigma \Delta_{(0)}\Delta_{(0)}\sigma+\frac{1}{3}h^2 \right]. 
 \end{eqnarray}
%
%
 Here $\Delta_{(1)}$ spin-1 Laplacian for transverse fields on $R^3$.
 We treat $h_{\tau i}, h^{\perp}_{~ij} , \zeta_{i}^{~\perp}, \sigma$ and $h$
as independent set of variables and substitute 
(\ref{expansgrav}) in (\ref{gravnorm}).  Performing the resultant integrals just as in the 
case of the $U(1)$ field we obtain 
the  change in measure as
\begin{equation}\label{grav-changem1}
 {\cal D} h_{ij}  = {\cal D} h_{ij}^\perp {\cal D}\zeta_i^\perp {\cal D} \sigma {\cal D} h  J, 
 \qquad  J = [{\rm det}\left(  ( -\Delta_{(1)}  ) ( -\Delta_{(0)})^2 \right) ]^{\frac{1}{2}}.
\end{equation}
Similarly, following the same steps for the Maxwell field  decomposing $h_{\tau i }$ into its transverse
and longitudinal component along $AdS_3$ results in the following change in measure
\begin{equation}\label{grav-changem2}
{\cal D} h_{\tau i } = {\cal D} h_{\tau i }^\perp {\cal D} \phi J, \qquad 
J = [{\rm det} ( -\Delta_{(0)}  ) ]^{\frac{1}{2} }.
\end{equation}

\subsection*{2-form  in $D=6$}

Again we start with the canonical normalization 
             \begin{equation}\label{normant}
\int {\cal D} A_{\mu\nu}  e^{-\int d^6 x \sqrt{g}  (A_{\mu\nu} A^{\mu\nu}) }=1.
\end{equation}
We decompose $A_{\mu\nu}$ in the components of $A_{\tau i}$ and $A_{ij}$.
where $i, j$ denotes the components along $AdS_5$ directions.
We can further decompose $A_{ij}$ in the following way
\begin{align}
    A_{ij}=A^{\perp}_{ij}+\nabla_i\chi^{\perp}_j-\nabla_j\chi^{\perp}_i, \qquad 
    \nabla_i A^{\perp ij}=0, \qquad  \nabla_i \chi^{\perp i}=0.
\end{align}
Note that the longitudinal component of $\chi^i$ does not occur in this decomposition. 
     Now 
             \begin{align}\label{antexpan}
                 \int d^6 x\sqrt{g} ( A_{\mu\nu}A^{\mu\nu}) 
                                             &=\int d^6 x\sqrt{g} \left[  2(A_{\tau i})^2+(A^{\perp}_{~ij})^2
                                             +2 \chi^{\perp}_i (-\Delta_{(1)} +   4) 
                                            \chi^{\perp\,  i} \right].
             \end{align}
             where $\Delta_{(1)} $ is the spin-1 laplacian on $AdS_5$.
             Substituting the expansion in (\ref{antexpan}) into ({\ref{normant}) and re-writing the integration 
             interms of $A^\perp_{ij},  \chi^\perp_i$ we obtain the following equation for the 
             change in measures
             \begin{equation}\label{2form-changem1}
             {\cal D} A_{ij} = {\cal D}{A}_{ij}^\perp {\cal D} \chi_i^\perp J, 
             \qquad J = [{\rm det} ( - \Delta_{(1)} + 4) ]^{\frac{1}{2} }.
             \end{equation}
             Similar to the $U(1)$ case, one can decompose the component $A_{\tau i }$  in terms of
             its transverse and longitudinal component along $AdS_3$. The change in measure for this 
             decomposition is given by 
             \begin{equation}\label{2form-changem2}
             {\cal D} A_{\tau i} = {\cal D} A_{\tau i}^\perp {\cal D} \chi_{\tau}  J, \qquad
             J = [ {\rm det} ( -\Delta_{(0)} ) ]^{\frac{1}{2} }.
             \end{equation}
  
  \subsection*{2-form gauge group: non-zero modes}
Let us also evaluate the change in measure for the gauge group integrations. 
We do this for the non-zero Kaluza-Klein modes first.  The reason we need these 
change of measures is because while gauge fixing, one divides out with the 
integrations over the measure ${\cal D} \epsilon_\mu$. However the gauge slice we choose
is over transformations which involve
 the transverse $\epsilon_i^\perp$, and longitudinal  component $ \epsilon$, on $AdS_5$ 
We first normalise  the gauge parameters $\epsilon_\mu$ as
\begin{equation} \label{mc1}
\int {\cal D} \epsilon_\mu  e^{ - \int d^6 x \sqrt{g} \epsilon_\mu \epsilon^\mu  }  = 1.
\end{equation}
In the gauge transformation for the $2$-form (\ref{antigt}), note that the transverse component 
$\epsilon_\mu^\perp$ over $S^1\times AdS_5$ occurs. Therefore 
we first make a change of variables by the transverse and longitudinal components of 
the gauge parameter on the entire $S^1\times AdS_5$ space.
For this we write
\begin{equation}
\epsilon_\mu = \epsilon_\mu^\perp + \nabla_\mu \tilde\epsilon, \qquad \mu \in \{ \tau,  i  \}.
\end{equation}
Then we write (\ref{mc1} ) as 
\begin{eqnarray}\label{mc2}
&& \int {\cal D} \epsilon_\mu  e^{ - \int d^6 x \sqrt{g} \epsilon_\mu \epsilon^\mu  } =
 \int {\cal D} \epsilon_\mu^\perp {\cal D} \tilde \epsilon  J_1
e^{ - \int d^6 x \sqrt{g}\left[   \epsilon_\mu^\perp \epsilon^{\perp \, \mu }   +
 \tilde \epsilon ( - \partial_\tau^2  - \Delta_{(0)})  \tilde\epsilon  \right]  } \\ \nonumber
& &  = \int {\cal D} \epsilon_\mu^\perp  J_1 [ {\rm det} ( -\partial_\tau^2 - \Delta_{(0)} ) ]^{-\frac{1}{2} }
 e^{  - \int d^6 x \sqrt{g}  \epsilon_\mu^\perp \epsilon^{\perp \, \mu }  }.
\end{eqnarray}
We can now further write $\epsilon_\mu^\perp$ in terms of $\epsilon $ and $\epsilon_i^\perp$ 
\begin{equation}
\{ \epsilon_\mu^\perp  \} \rightarrow \{ \epsilon, \epsilon_i^\perp\}, \qquad 
\nabla^i \epsilon_i^\perp =0.
\end{equation}
Note that here we need to eliminate $\epsilon_\tau $ in terms of $\epsilon$ using equation
(\ref{anticonstraint}). 
Proceeding  we write the integral as 
\begin{eqnarray}
\int d^6 x \sqrt{g}(  \epsilon_\mu^\perp  \epsilon^{\perp \, \mu }  )  &=& \int d^6x \sqrt{g}  \left[ 
 \epsilon\left( \frac{ -\Delta_{(0)}^2  }{\partial_\tau^2} 
\right)\epsilon + \epsilon_i^\perp \epsilon^{\perp i } + 
\epsilon ( -\Delta_{(0)}) \epsilon  \right], \\ \nonumber
&=&  \int d^6x \sqrt{g}  \left[ 
\epsilon_i^\perp \epsilon^{\perp i }  + \epsilon \left( \frac{1}{\partial_\tau^2} [
-\partial_\tau^2 - \Delta_{(0)} ]( -\Delta_{(0)} )  \right )  \epsilon \right].
\end{eqnarray}
To obtain the  first line we have eliminated $\epsilon_\tau$ using ( \ref{anticonstraint}).
Now substituting the above equation in the exponential  of (\ref{mc2}), we obtain 
\begin{eqnarray}\label{mc3}
 & & \int {\cal D} \epsilon_\mu^\perp  J_1 [ {\rm det} ( -\partial_\tau^2 - \Delta_{(0)} ) ]^{-\frac{1}{2} }
 e^{  - \int d^6 x \sqrt{g}  \epsilon_\mu^\perp \epsilon^{\perp \, \mu }  } \\ \nonumber
 &=&
 \int {\cal D} \epsilon_i^\perp{\cal D} \epsilon J_1 J_2  [ {\rm det} ( -\partial_\tau^2 - \Delta_{(0)} ) ]^{-\frac{1}{2} }
 e^{  - \int d^6 x \sqrt{g}\left[   \epsilon_i^\perp \epsilon^{\perp i }  + \epsilon \left( \frac{1}{\partial_\tau^2} ( 
-\partial_\tau^2 - \Delta_{(0)} )( -\Delta_{(0)} )  \right )  \epsilon \right] } , \\ \nonumber
&=&  J_1J_2  \left[ {\rm det} \left( \left( \frac{1}{\partial_\tau} \right) 
( -\partial_\tau^2 - \Delta_{(0)} )  \right) \right]^{-1} [{\rm det} ( - \Delta_{(0)} ) ]^{-\frac{1}{2}} .
\end{eqnarray}
Using (\ref{mc1}) and (\ref{mc2}) and (\ref{mc3}) , we obtain  the following 
change in measure 
\begin{eqnarray}\label{gauge-changem1}
{\cal D}\epsilon_\mu  &=& 
 {\cal D} \epsilon_i^\perp {\cal D} \epsilon {\cal D} \tilde \epsilon   J, 
\\ \nonumber
J  &=& J_1J_2   =  \left[ {\rm det} \left( \left( \frac{1}{\partial_\tau} \right) 
( -\partial_\tau^2 - \Delta_{(0)} )  \right) \right] [{\rm det} ( - \Delta_{(0)} ) ]^{\frac{1}{2}} .
\end{eqnarray}

\subsection*{2-form gauge group: zero modes}

Now let us find the change in the measure of the gauge parameter for the zero modes. 
Again we begin with the normalization in (\ref{mc1}).  For the zero modes 
The constraint $\nabla^\mu \epsilon_\mu=0$ reduces to the constraint 
$\nabla^i \epsilon_i =0$, so we 
 write $\epsilon_i  = \epsilon_i^\perp + \nabla_i \epsilon$. Then 
\begin{equation}
\int d^6 x \sqrt{g} \epsilon_\mu \epsilon^\mu = 
\int d^6 x \sqrt{g}  ( \epsilon_\tau ^2 + \epsilon_i^\perp \epsilon^{\perp \, i } 
+ \epsilon ( -\Delta_0) \epsilon  ) .
\end{equation}
We substitute this into (\ref{mc1}) and obtain
\begin{eqnarray}\label{mc4}
\int {\cal D} \epsilon_\mu  e^{ - \int d^6 x \sqrt{g} \epsilon_\mu \epsilon^\mu  } 
 &=&\int {\cal D} \epsilon_\tau {\cal D} \epsilon_i^\perp {\cal D} \epsilon J 
e^{ \int d^6 x \sqrt{g}  ( \epsilon_\tau ^2 + \epsilon_i^\perp \epsilon^{\perp \, i } 
+ \epsilon ( -\Delta_0) \epsilon  )  },  \\ \nonumber
& =& J [{\rm det} ( -\Delta_0)]^{-\frac{1}{2} } =1. 
\end{eqnarray}
Now comparing (\ref{mc4}) and (\ref{mc1}) we obtain 
\begin{eqnarray} \label{gauge-changem2}
 {\cal D} \epsilon_\mu  =  {\cal D} \epsilon_\tau {\cal D} \epsilon_i^\perp {\cal D}\epsilon J  \qquad \qquad
J = [{\rm det} ( -\Delta_0)]^{\frac{1}{2} }.
\end{eqnarray}

\section{Partition function of the graviton: $S^1\times R^3$} \label{appen4}

Consider the action of the graviton in flat space. 
This is given by 
\begin{eqnarray}
S=  \int d^4x \left[ - \partial_\mu h^{\mu \nu} \partial_\alpha h^\alpha_{\; \nu} 
+ \frac{1}{2} \partial^\alpha h_{\mu\nu} \partial_\alpha h^{\mu\nu}
+ \partial_\mu h^{\mu\nu} \partial_\nu h^\alpha{\;\alpha}
- \frac{1}{2} \partial_\alpha h^\mu_{\;\; \mu} \partial^\alpha h^\nu_{\;\; \nu}
\right]. 
 \nonumber \\
\end{eqnarray}
To evaluate the partition function on $S^1\times R^3$ we gauge fix as done for the Maxwell field 
and the $2$-form in $D=6$. 
We first write the action in terms of $h_{\tau\tau}, h_{\tau i}, h_{ij}$. 
This is given by the following 
\begin{eqnarray}
S &=& \int d^4 x (  {\cal L}_1 + {\cal L}_2 + {\cal L}_3) , \\ \nonumber
{\cal L}_1 &=& h^{\tau j} ( - \partial^i \partial_i ) h_{\tau j } - ( \partial_i h^{i \tau} ) ^2, 
\\ \nonumber
{\cal L}_2 &=& h \partial^i \partial_i h_{\tau\tau} + \partial^i h_{ij} \partial^j h_{\tau\tau} 
+ 2 \partial_\tau h_{\tau i } ( \partial^i h - \partial_j h^{ij} ) , 
\\ \nonumber
{\cal L}_3 &=& \frac{1}{2} h^{ij} ( -\partial_\tau^2 - \partial^k \partial_k ) h_{ij}
- \frac{1}{2} h ( - \partial_\tau^2  - \partial^k \partial_k ) h 
- \partial_i h^{ij} \partial^k h_{kj} 
+ \partial_i h^{ij} \partial_j h . 
\end{eqnarray}
This action is invariant under
\begin{eqnarray} \label{gaugetrans2}
\delta h_{\tau\tau} &=& 2\partial_\tau \epsilon_\tau, \\ \nonumber
\delta h_{\tau i } = \partial_\tau \epsilon_i + \partial_i \epsilon_\tau, \\ \nonumber
\delta h_{i j} = \partial_i \epsilon_j  + \partial_j \epsilon_i. 
\end{eqnarray}
We  decompose
\begin{eqnarray}
h_{ij} &=& \bar h_{ij} + \frac{1}{3} \delta_{ij} h ,  \\ \nonumber
\bar h_{ij} &=& h_{ij}^\perp + \partial_i \zeta_j^\perp + \partial_j \zeta_i^\perp
+ \partial_{i} \partial_j \sigma -\frac{1}{3} \delta_{ij} \sigma, \\ \nonumber
& & {\rm where} \; \partial^i h_{ij}^\perp =0, \qquad \partial^i \zeta_i^\perp =0.
\end{eqnarray}
Under this change of variables the measure changes as  (\ref{grav-changem1}) 
\begin{eqnarray}\label{g1}
{\cal D} h_{\tau\tau} {\cal D} h_{\tau i } {\cal D} h_{ij} &= &
{\cal D} h_{\tau\tau} {\cal D} h_{\tau i } {\cal D} h_{ij}^\perp {\cal D} \zeta_i^\perp 
{\cal D} h {\cal D} \sigma \times J , \\ \nonumber
J &=& {\rm det} [ (  - \Delta_{(1)} ) ( -\Delta_{(0)} ) ^2]^{\frac{1}{2}}
\end{eqnarray}
Here we have used the subscript on the Laplacian's to denote that act on transverse vectors and 
scalars respectively. 
The measure can be split for the zero and the non-zero Kaluza Klein modes on $S^1$ as
\begin{eqnarray}
{\cal D} h_{\tau\tau} {\cal D} h_{\tau i } {\cal D} h_{ij} &= &
\left[ {\cal D} h_{\tau\tau} {\cal D} h_{\tau i } {\cal D} h_{ij}^\perp {\cal D} \zeta_i^\perp 
{\cal D} h {\cal D} \sigma \times J \right]_{n\neq 0}  \\ \nonumber
&& \times  {\cal D} h_{\tau\tau} {\cal D} h_{\tau i } {\cal D} h_{ij}^\perp {\cal D} \zeta_i^\perp 
{\cal D} 
\left[ {\cal D} h {\cal D} \sigma \times J \right]_{n =0} . 
\end{eqnarray}

\subsection*{Non-zero Kaluza-Klein modes}

For the  non-zero Kaluza-Klein modes ,  from the gauge transformation  in (\ref{gaugetrans2}) 
we can fix the gauge 
\begin{eqnarray}
h_{\tau\tau} =0, \qquad h_{\tau i } =0. 
\end{eqnarray}
This  gauge choice along with the Fadeev-Popov determinant  is given by 
\begin{eqnarray}\label{g2}
{\cal G} = \delta ( h_{\tau \tau} ) ( h_{\tau i}) [ {\rm det} ( \partial_\tau)^4]. 
\end{eqnarray}
Imposing the delta function constraints results 
 in the following action
\begin{eqnarray}
S &=& \int d^4 x ( {\cal L}_{h^\perp}+  {\cal L}_{\zeta^\perp}+ {\cal L}^\prime_3 )_{\sigma, h}, 
\\ \nonumber
{\cal L}_{h^\perp} &=& \frac{1}{2} h^{ij \perp} ( -\partial_\tau^2 - \Box) h_{ij}^\perp, 
\\ \nonumber
{\cal L}_{\zeta^\perp}&=& \zeta^{i \perp} ( - \partial_\tau^2 ) ( -\Box) \zeta_i^\perp, 
\\ \nonumber
{\cal L}_{\sigma, h} &=& 
\frac{1}{9} 
\left[  \sigma ( - 3 \partial_\tau^2  + \Box) \Box^2 \sigma 
- 2 h \Box^2 \sigma + h ( 3 \partial_\tau^2  + \Box ) h  \right]. 
\end{eqnarray}
It is convenient to write the $\sigma, h$ part of the action as a matrix.  
\begin{eqnarray}
{\cal L}_{\sigma, h}  = \frac{1}{9} \left( \begin{array}{c c} 
\sigma  & h  \end{array} \right) 
\left(  \begin{array}{ c c}
 ( - 3 \partial_\tau^2  + \Box) \Box^2 & 
-  \Box^2  \\
-\Box^2 & 
 ( 3 \partial_\tau^2  + \Box )   
\end{array} \right)
\left( \begin{array}{c}
\sigma \\ h 
\end{array}
\right) . 
\end{eqnarray}
Integrating out $h_{ij}^\perp, \zeta_i^\perp, \sigma, h$ we obtain
\begin{eqnarray} \label{g3}
{\cal Z}_{h_{ij}^\perp, \zeta_i^\perp, \sigma, h} = [{\rm det}  ( -\partial_\tau^2 - \Delta_{(2)} ) 
( -\Delta_{(1)} )( -\Delta_{(0)} )^2 ( \partial_\tau^8) ]^{-\frac{1}{2}}
\end{eqnarray}
Here we have used the fact that $\zeta_i^\perp$  has 2 independent degrees of freedom, therefore
${\rm det}(  \partial_\tau^2 ( -\Delta_{(1)} ) )  = 
(  {\rm det} ( \partial_\tau)  )^4 {\rm det} ( -\Delta_{(1)} ) $ \footnote{The easiest way to see this 
is to work in momentum space and use the projector $ \delta_{ij} - \frac{k_i k_i}{k^2}$ 
to ensure $\partial \zeta_i^\perp = 0$}. 
Now combining (\ref{g3}), the Jacobian  in (\ref{g1}) and the Fadeev-Popov determinant (\ref{g3}) 
we obtain 
\begin{eqnarray} \label{n1}
{\cal Z}_{n\neq0} = [ {\rm det}  ( -\partial_\tau^2 - \Delta_{(2)} )]^{-\frac{1}{2}}. 
\end{eqnarray}

\subsection*{Zero modes}

For the Kaluza-Klein zero modes,  we see that $h_{\tau\tau}$ is gauge invariant, and 
we have 
\begin{eqnarray} \label{gaugetrans3}
\delta h_{\tau\tau} &=&0, \qquad \qquad 
\delta h_{\tau i } =  \partial_i \epsilon_\tau, \\ \nonumber
\delta h_{i j} &=& \partial_i \epsilon_j  + \partial_j \epsilon_i. 
\end{eqnarray}
Now we decompose the fields as 
\begin{eqnarray}
h_{\tau i} &=& h_{\tau i}^\perp + \partial_i \phi, \qquad  \partial^i h_{\tau i }^\perp = 0 , \\ \nonumber
h_{ij} &=& \bar h_{ij} + \frac{1}{3} \delta_{ij} h ,  \\ \nonumber
\bar h_{ij} &=& h_{ij}^\perp + \partial_i \zeta_j^\perp + \partial_j \zeta_i^\perp
+ \partial_{i} \partial_j \sigma -\frac{1}{3} \delta_{ij} \sigma, \\ \nonumber
 {\rm where}  &&\; \partial^i h_{ij}^\perp =0, \qquad \partial^i \zeta_i^\perp =0. 
\end{eqnarray}
    This change of variable involves the Jacobian (\ref{grav-changem1}), ( \ref{grav-changem2})
\begin{eqnarray}\label{f3}
{\cal D} h_{\tau \tau} {\cal D} h_{\tau i } {\cal D} h_{ij} 
&=& {\cal D} h_{\tau\tau} {\cal D} h_{\tau i}^\perp {\cal D} \phi {\cal D} h_{ij}^\perp {\cal D} \zeta^\perp_i {\cal D} \sigma {\cal D} h \times J', \qquad \\ \nonumber
J' &=& {\rm det} [ ( - \Delta_{(1)} ) ( -\Delta_{(0)}) ^3]^{\frac{1}{2}}. 
\end{eqnarray}
Decomposing the gauge parameter we have
\begin{eqnarray}
\epsilon_i = \epsilon_i^\perp  + \frac{1}{2}{\partial_i} \epsilon. 
\end{eqnarray}
In terms of the components, the gauge transformation can be written as
\begin{eqnarray} \label{zogag}
\delta h_{ij}^\perp = 0, \quad \delta h_{\tau i}^\perp = 0, \quad
\delta \phi = \epsilon_\tau, \quad \delta \zeta_i^\perp =\epsilon_i^\perp, \quad
\delta \sigma = \epsilon, \qquad \delta h = \Box\epsilon
\end{eqnarray}
The action in terms of the components becomes
\begin{eqnarray}
S &=& \int d^4x \left( {\cal L }_{h_{\tau\tau} } + {\cal L}_{h_{\tau i }^\perp} + 
{\cal L}_{h_{ij}^{\perp} }+ {\cal L }_{\sigma, h} \right) , \\ \nonumber
{\cal L}_{h_{\tau\tau}} &=& \frac{2}{3} ( h - \Box \sigma) \Box h_{\tau\tau} , \\ \nonumber
{\cal L}_{h_{\tau i }^\perp} &=& h^{\tau i \perp} ( -\Box) h_{\tau i }^\perp, \\ \nonumber
{\cal L }_{h_{ij}^\perp} &=& \frac{1}{2} h^{ij \perp} ( - \Box) h_{ij}^\perp, \\ \nonumber
{\cal L }_{\sigma, h} &=&\frac{1}{9} \left( \begin{array}{c c} 
\sigma  & h  \end{array} \right) 
\left(  \begin{array}{ c c}
  \Box^3 & 
-  \Box^2  \\
-\Box^2 & 
 \Box    
\end{array} \right)
\left( \begin{array}{c}
\sigma \\ h 
\end{array}
\right) . 
\end{eqnarray}
Note the absence of $\phi, \zeta_i^\perp$ in the action due to gauge invariance. 
We fix gauge by introducing 
\begin{eqnarray}
{\cal G} = \delta(\phi) \delta( \zeta_i^\perp) \delta( \sigma) .
\end{eqnarray}
It is clear from the gauge transformations given in (\ref{zogag}), the Fadeev-Popov determinant is unity. 
Since our gauge parameters are $\epsilon_i^\perp, \epsilon$ we need the following change 
of variables 
\begin{eqnarray} \label{f1}
{\cal D} \epsilon_i^\perp {\cal D} \epsilon = {\cal D} \epsilon_i \times J'', 
\qquad J'' = \left[ \frac{1}{ \Delta_{(0)} }  \right]^{\frac{1}{2} }.
\end{eqnarray}
Now we can integrate over $\epsilon_\tau, \zeta_i^\perp, \sigma$, which implements the 
delta function. 
After which we can integrate over $h_{\tau, i}^\perp, h_{ij}^\perp, h$ and finally $h_{\tau\tau}$. 
This results in 
\begin{eqnarray}\label{f2}
{\cal Z}_{h_{ij}^\perp, h_{\tau i }^\perp, h, \sigma, \zeta_i^\perp,  h_{\tau\tau} }
= \{ {\rm det} [( -\Delta_{(2)} ) ( -\Delta_{(1)} ) ( -\Delta_{( 0) } )^2] \}^{\frac{1}{2}}.
\end{eqnarray}
Now putting together the Jacobian  from the change of measure in (\ref{f3}), 
and that for the gauge transformation in (\ref{f1}) along with the result of the integrations
in (\ref{f2}) we obtain the final result for the zero modes partition function to be given by 
\begin{equation}\label{n0}
{\cal Z}_{n =0} =\left[  \frac{1}{ {\rm det} ( -\Delta_{(2)} ) }  \right]^{\frac{1}{2}}. 
\end{equation}

Combining the result in (\ref{n0}) and (\ref{n1}) we obtain the following partition function for the 
graviton on $S^1 \times R^3$
\begin{equation}
{\cal Z} = [ {\rm det} ( - \partial_\tau^2 - \Delta_{(2)} ) ]^{-\frac{1}{2} }.
\end{equation}
where $-\partial_\tau^2  \rightarrow n^2/q^2$ for each Kaluza-Klein mode and 
$\Delta_{(2)}$ is the Laplacian on transverse symmetry traceless tensor on $R^3$. The partition 
function can also be written as 
\begin{eqnarray}
{\cal Z} = \int {\cal D} h_{ij}^\perp \exp\left[
-\frac{1}{2}  \int d^4 x  \; h^{ij \; \perp} ( - \partial_\tau^2 - \partial^k \partial_k ) h_{ij}^\perp
\right].
\end{eqnarray}

\section{Brietenholer-Freedman bound  for higher spins} \label{appen5}

The equations of motions of massive symmetric traceless tensors in $AdS_{d}$
\footnote{Our signature for Lorentzian $AdS_{d-1} $ is $( -1, 1, 1, \cdots)$, and the radius of 
$AdS$ is chosen to be unity.} dimensions 
are given by  \cite{Metsaev:2003cu,Buchbinder:2006ge}
\begin{eqnarray}
( \nabla^2 - m_s^2 ) \Phi_{\mu_1 \mu_1 \cdots \mu_s} = 0,\\ \nonumber
\nabla^\mu \Phi_{\mu \mu_2 \cdots \mu_s}  =0, \\ \nonumber
g^{\mu\nu} \Phi_{\mu\nu\mu_3\cdots \mu_s} =0.
\end{eqnarray}
Here $\Phi_{\mu_1, \mu_2, \mu_s}$ is a totally symmetric rank $s$ tensor. 
The mass $m_s^2$ is given by 
\begin{equation}\label{defmass}
m_s^2 =  [ s^2 + s( d-6) + 6 -2d] + m^2 ,
\end{equation}
The term in the square brackets is the natural mass that exists due to the curvature of $AdS_d$. 
Note that it also exists when $m=0$, in fact `massless' higher spin  fields in $AdS_d$
have $m=0$. 
The  putative dual operator corresponding to the higher spin field by the 
$AdS_d/CFT_{d-1}$ corresponds to an operator of spin $s$ with conformal dimension
\cite{Metsaev:2003cu}
\begin{equation}
\Delta =  \frac{d-1}{2} + \sqrt{ m^2  + ( s + \frac{d-5}{2} )^2  }.
\end{equation}
For the reader who is familiar with this dictionary for scalars, note that taking $s=0$, the 
expression for the scaling dimension reduces to  
\begin{equation}\label{deltaspin}
\Delta = \frac{d-1}{2} + \sqrt{ m_{s=0}^2 + ( \frac{d-1}{2} ) ^2 }.
\end{equation}
This is the standard formula for the conformal dimension of the operator dual to the 
minimally coupled scalar of mass $m_{s=0}^2$ in $AdS_{d}$. 

Now let us consider the case when the conformal dimension is $\Delta = \frac{d-1}{2}$ or the 
term in the square root of (\ref{deltaspin}) vanishes, 
we have
\begin{equation} \label{bfbound}
m^2|_{\rm BF}  = - ( s + \frac{d-5}{2} )^2.
\end{equation}
The mass then satisfies the Breitenlohner-Freedman bound for the spin $s$ field in $AdS_d$. 
Using 
(\ref{bfbound}) in (\ref{defmass}) we obtain
\begin{equation} \label{bfbound2} 
m_s^2 |_{\rm BF}= - s - (  \frac{d-1}{2} )^2.
\end{equation}
Note that for the scalars this condition reduces to the familiar condition \cite{Breitenlohner:1982jf}
\begin{equation}
m^2_{s=0}|_{\rm BF} = - ( \frac{d-1}{2} )^2.
\end{equation}

The eigenvalue of the spin-$s$ Laplacian  $\Delta_{(s)}$ on Euclidean  $AdS_d$ is given by  \cite{Camporesi:1994ga}
\begin{equation}
\Delta_{(s)}\psi^{(\lambda, u )}_{\mu_1, \mu_2 \cdots \mu_s} =-  (  \lambda^2 + s + (  \frac{d-1}{2} )^2) 
 \psi^{(\lambda)}_{\mu_1, \mu_2 \cdots \mu_s} ,
\end{equation}
where $ \psi^{(\lambda, u)}_{\mu_1, \mu_2 \cdots \mu_s} $ are the corresponding eigenfunctions, 
$u$ are other quantum numbers characterizing the eigenfunctions. 
Therefore the eigenvalues of   $\Delta_{(s)} - m_s^2$ is  given by $ -\lambda^2$. 
\begin{equation} \label{eigenspins}
(\Delta_{(s)}  - m_s^2)  \psi^{(\lambda, u)}_{\mu_1, \mu_2 \cdots \mu_s}  = - \lambda^2  \psi^{(\lambda, u)}_{\mu_1, \mu_2 \cdots \mu_s} .
\end{equation}
Note that once we demand that  mass of the higher spin field saturates its Breitenlohner-Freedman 
bound,  the eigenvalues of the corresponding massive higher spin 
field is independent of the spin and is given by $-\lambda^2$.

\section{Twist weight from the replica cone: Maxwell field } \label{appen2}

In this section we review the evaluation of the expectation value of the Maxwell  stress tensor 
on the replica cone in $D=4$.  This was originally done in \cite{Candelas:1977zza}, we adapt it so that the method
can be extended for the 2-form in $D=6$. 
The geometry  is same as that discussed in section \ref{sect:2form}  for the planar  co-dimension 2 defect. 
To evaluate the expectation value of the stress tensor we  use the replica trick and consider the 
theory  on the cone. The transverse directions to the defect 
  are labelled as $x^a$ with $a \in \{1, 2\}$ and the 
 longitudinal directions are labelled as $x^i$ with $ i \in \{3, 4 \}$. 
Stress tensor for  $U(1)$ theory  is given by
      \begin{align}
      T_{\mu\nu} &
      =F_{\mu}^{~\lambda}F_{\nu\lambda}-\frac{1}{4}g_{\mu\nu}F_{\alpha\beta}F^{\alpha\beta},
      \\ \nonumber
      F_{\mu\nu} & = \partial_\mu A_\nu - \partial_\nu A_\mu.
      \end{align}
      The $U(1)$ theory is gauge invariant under the transformation
      \begin{equation}
      A_\mu \rightarrow A_\mu + \partial_\mu \epsilon.
      \end{equation}
      Using this symmetry we can fix the  covariant gauge  
      \begin{equation}\label{covcond}
      \partial^\mu A_\mu =0.
      \end{equation}
      The equations of motion in this gauge reduces to 
      \begin{equation}\label{maxeom}
      \nabla^2 A_\mu =0.
      \end{equation}
      The still allows us the transformation 
      \begin{equation}
      A_\mu' = A_\mu + \partial_\mu \epsilon, \qquad  {\rm with }\quad \Box \epsilon =0.
      \end{equation}
      Then given a gauge potential which satisfies (\ref{covcond}) and (\ref{maxeom}) we can make 
      a further gauge transformation  so that 
      \begin{equation}\label{maxgcond}
      \partial^a A_a' =0, \qquad \partial^i A_{i}' =,0  \quad a \in \{ 1, 2\} , i \in \{3, 4\}.
      \end{equation}
      This can be achieved by choosing the gauge  transformation  to be
      \begin{equation}
      \epsilon  = - \frac{\partial^i A_i }{\widehat\nabla^2} , \qquad \widehat\nabla^2 = \partial_3^2 + \partial_4^2.
      \end{equation}
      Note that the gauge transformation also satisfies $\Box \epsilon=0$. 
      
      We need to evaluate the expectation value of the stress tensor on the replica cone. 
      For this it is convenient to choose polar coordinates
      \begin{equation}
      ds^2 = dr^2  + r^2 d\theta^2 + ( dx^3)^2 + ( dx^4)^2.
      \end{equation}
      Here $\theta \sim \theta + 2\pi q $. 
      The two point function fo the gauge field  on the cone satisfying the gauge condition 
      (\ref{maxgcond}), is given by 
      \begin{eqnarray}
      \tilde G_{\mu\nu'} ( x,x')&=& \langle A_\mu ( x) A_{\nu'} ( x') \rangle.   \\ \nonumber
      \tilde G_{ab'} (x,x') &=&  - \frac{P_a P_{a'} }{ \widehat\nabla^2} \tilde G (x, x'),  \qquad
      \tilde 
      G_{i j'}(x, x')    = - \left[ \delta_{ij}  - \frac{\partial_i\partial_j}{ \widehat\nabla^2} \right] \tilde G(x, x') ,
      \\ \nonumber
      \tilde G_{a i'} (x,x') &=& \tilde G_{i a'} ( x, x') =0.
      \end{eqnarray}
      where $\tilde G(x, x') $ is the scalar propagator on the cone which is given by 
      \begin{eqnarray}
      \tilde G(x, x') &=& \frac{1}{ 4q\pi^2 r r' ( a - a^{-1} )} \frac{ a^{\frac{1}{q} }- a^{- \frac{1}{q}} }{
      a^{\frac{1}{q} }+  a^{- \frac{1}{q}} - 2 \cos \left( \frac{\theta - \theta'}{q} \right) }, \\ \nonumber
      \frac{a}{ 1+ a^2} &=& \frac{rr'}{ ( x^ i - x^{\prime i } )^2 + r^2 + r^{\prime 2} }.
      \end{eqnarray}
      and $P_a$ is defined as in (\ref{defpa}). 
      To remove the divergences that occur in the point split approach for evaluating the 
      expectation value of the stress tensor we regulate the Greens function by considering
      \begin{equation} \label{reggreen}
      G_{\mu\nu'} ( x, x')  = \tilde G_{\mu \nu'} ( x, x') |_{q} - \tilde G_{\mu \nu'} ( x, x') |_{q=1}, 
      \qquad 
      G( x, x')  = \tilde G ( x, x') |_{q} - \tilde G_( x, x') |_{q=1}.
      \end{equation}
      The expectation value of the stress tensor is then obtained by 
     taking the following limit
     \begin{equation}
    \langle  T_{\mu\nu}(x) \sigma_q \rangle  = \lim_{x\rightarrow x'} 
     \left( \langle F_{\mu \lambda} ( x) F_{\nu}^{\; \lambda'} (x')  \rangle 
     -\frac{g_{\mu\nu} }{4} \langle F_{\rho\sigma}(x) F^{\rho'\sigma'} ( x') \rangle \right) .
     \end{equation}
     Here the two point function on the RHS of the above equation are obtained by 
     using the regularised Greens function in (\ref{reggreen}). 
     This leads to $6$ terms in the following equation
     \begin{eqnarray}\label{fullmaxstress}
     \langle  T_{\mu\nu}(x) \sigma_q \rangle  
     &=& \lim_{x\rightarrow x'} \left[ 
     \nabla_{\mu\nu'}G_{\lambda}^{~\lambda'} (x, x') 
     -\nabla_{\mu}^{~\lambda'}G_{\lambda\nu'}(x, x') 
     -\nabla_{\nu'}^{~\lambda}G_{\mu\lambda'}(x, x') +\nabla_{\lambda}^{\lambda'}G_{\mu\nu'}(x, x') 
     \right. 
     \nonumber \\
     & & \left. 
     -\frac{1}{2}g_{\mu\nu'}(x) (\nabla_{\mu}^{\mu'}G_{\nu}^{\; \nu'} (x, x') 
     -\nabla_{\mu}^{\nu'}G_{\nu}^{\;\mu'} ( x, x') ) \right].
     \end{eqnarray}
     where $\nabla_{\mu\nu'} = \nabla_\mu \nabla_{\nu'}$. 
     
      To proceed further we would need the following relation
      \begin{equation}\label{rel3}
      \lim_{x\rightarrow x'} P_a P^{a\prime}  G(x, x') 
      = \lim_{x\rightarrow x'}\nabla_a \nabla^{a'}  G(x, x') .
      \end{equation}
      The repeated indices always mean that we sum over them, inspite of the fact that 
      one  set of indices are labelled with a prime. 
      We also have 
      \begin{eqnarray} \label{rel1}
       (  \nabla_1 P_2-\nabla_2 P_1) G ( x, x')
                                    &=&
                                    -r\left(
                                    \frac{\partial^2}{ \partial r^2}
                                    + \frac{1}{r} \frac{\partial}{  \partial r}+ \frac{1}{r^2} \frac{\partial^2 }{\partial \theta^2}\right)G(x, x'),\nonumber\\
                                    &=&r\widehat\nabla^2 G(x, x') .
      \end{eqnarray}
      In the last line we have used the equation satisfied by the Scalar Greens function for 
      when $x\neq x'$; 
       $\Box G(x, x')  =0$. 
       Finally 
       \begin{eqnarray} \label{rel2}
        \nabla^{2'} P^{1'}-\nabla^{1'}P^{2'}G(x, x') = -\frac{1}{r'}\widehat\nabla^2 G(x, x').
        \end{eqnarray}
        To arrive at this result we need to use the Christofel symbols of the polar coordinates.
        
Let us now analyse the stress tensor in (\ref{fullmaxstress}) term by term. 
We first study when the external indices are both longitudinal direction to the defect.
The first term is given by
\begin{eqnarray} \label{1term}
&&\nabla_{ij'}G_{\lambda}^{\;\lambda'} (x, x') =   \nabla_{ij'}(  G_{k}^{~k'}+G_{c}^{~c'})\nonumber\\
                                          &=& \nabla_{ij'}\left[-\left(\delta_{k}^{k}-\frac{\partial_k\partial^k}
                                          {\widehat\nabla^2}\right)-\frac{P_c P^{c'}}{\widehat\nabla^2}\right]G (x, x') 
                                          = -\left( 1+\frac{P_c P^{c'}}{\widehat\nabla^2}\right) \nabla_{ij'} 
                                          {G} (x, x'), \nonumber\\
                                          &=&\frac{1}{2}\delta_{ij}[\widehat\nabla^2+P_aP^{a'}]{G}(x, x') .
\end{eqnarray}
To arrive at the line of the  above
 equation we have neglected terms which vanish on taking the $x\rightarrow x'$ limit. 
More explicitly it can be seen that  the following identities  hold
\begin{eqnarray}\label{limitindent}
\lim_{x\rightarrow x'} (\nabla^2  ) \left( \nabla_{ij' }  + \frac{\delta_{ij}}{2} \widehat\nabla^2 \right)G(x, x')  &=&0,
\\ \nonumber
\lim_{x\rightarrow x'} (P_cP^{c'}) \left( \nabla_{ij' }  + \frac{\delta_{ij} }{2} \widehat\nabla^2 \right)G(x, x')  
&=&0, 
\\ \nonumber
\lim_{x\rightarrow x'} (
\nabla_c\nabla^{c'}) \left(  \nabla_{ij' }  + \frac{\delta_{ij}}{2} \widehat\nabla^2 \right)G(x, x')
&=&0.
\end{eqnarray}
The second and the third term  in (\ref{fullmaxstress}) 
vanish since the Greens function satisfies the gauge condition
(\ref{maxgcond}) 
For example consider the second term
\begin{eqnarray}\label{2term}
-\nabla_i^{\lambda \prime} G_{\lambda j'} (x, x') = - \nabla_i \nabla^{k'} G_{k j'} (x, x')  =
+ \nabla_i \nabla^{k} G_{k j'}(x, x')   =0.
\end{eqnarray}
We have converted the derivative on the primed coordinate to the unprimed 
using translational invariance of the Greens function in the longitudinal directions. 
Similarly the third term in (\ref{fullmaxstress}) vanishes
\begin{equation}\label{3term}
-\nabla_{j'}^{\lambda}G_{i \lambda'} (x, x') =0.
\end{equation}
Let us now look at the fourth term
\begin{eqnarray} \label{4term}
  +\nabla_{\lambda}^{\lambda'}G_{ij'}(x, x') &=& -(\nabla_k\nabla^{k'}+\nabla_c\nabla^{c'})
  \left( \delta_{ij}-\frac{\partial_i\partial_j}{\widehat\nabla^2} \right) G (x, x') ,     \nonumber\\
                                         &=&-(-\widehat\nabla^2+\nabla_c\nabla^{c'})\left[ \frac{1}{2}\delta_{ij} \right] G(x, x') .
\end{eqnarray}
In the last line we have used the identities in (\ref{limitindent}) to  simplify the  term proportional to 
$\frac{\partial_i\partial_j}{\widehat\nabla^2}$. 
Now let us look the fifth term in (\ref{fullmaxstress}). 
\begin{align}\label{5term}
        -\frac{1}{2}\delta_{ij}(\nabla_{\mu}^{\; \mu'}G_{\nu}^{\nu'})&=-\frac{1}{2}\delta_{ij}  (\nabla_k\nabla^{k'}+\nabla_c\nabla^{c'})
        \left[-(\delta_{k}^{k}-\frac{\partial_k\partial^k}{\widehat\nabla^2})-\frac{P_d P^{d'}}{\widehat\nabla^2}
        \right]{G}(x, x'),  \nonumber\\
                                                                             &=\frac{1}{2}\delta_{ij} (-\nabla^2+\nabla_c\nabla^{c'})(1+\frac{P_d P^{d'}}{\widehat\nabla^2})\tilde{G}(x, x'), \nonumber\\
                                                                             &=\frac{1}{2}\delta_{ij}\left[-\widehat\nabla^2+\frac{\nabla_c P_{d}\nabla^{c'}P^{d'}}{\widehat\nabla^2}\right]{G}(x, x') .
       \end{align}     
Finally the sixth term  is given by 
\begin{eqnarray}\label{6term}
     \frac{1}{2}\delta_{ij} \nabla_{\mu}^{\nu'}G_{\nu}^{\mu} (x, x') 
     &=& \frac{1}{2}\delta_{ij} \nabla_{c}^{d'}G_{d}^{c'}(x, x'), \nonumber\\
                                                                           &=&-\frac{1}{2}\delta_{ij}\frac{\nabla_c P_{d}\nabla^{d'}P^{c'}}{\widehat\nabla^2}{G}(x,x') .
\end{eqnarray}
To arrive at the first equality in the above equation we have used the fact that Green's function 
satisfies the gauge condition $\nabla_i^{l'} G_l^{k'} =0$. 
Let us now combine the fifth  and the sixth term in (\ref{5term}) and (\ref{6term}). 
We obtain  
\begin{eqnarray}\label{65term}
     &&    \frac{1}{2}\delta_{ij}\left[-\widehat\nabla^2+  \frac{\nabla_c P_{d}(\nabla^{c'}P^{d'}-\nabla^{d'}P^{c'})}{\widehat\nabla^2}\right]{G}(x, x' ), \\ \nonumber
     &&    =\frac{1}{2}\delta_{ij}\left[-\widehat\nabla^2+\frac{(\nabla_1P_2-\nabla_2P_1)(\nabla^{1'}P^{2'}-\nabla^{2'}P^{1'}}{\widehat\nabla^2})\right]{G} (x, x') ,\nonumber\\ \nonumber
                                    &&= \frac{1}{2}\delta_{ij}[-\widehat\nabla^2+\widehat\nabla^2]{G}(x, x')  =0.
                                 \end{eqnarray}
The last line is obtained by using the identities in (\ref{rel1}) and (\ref{rel2}) and also set $r=r'$ external 
to the differential operators. 
We have shown that adding the 5th and the 6th term vanishes.
From the origin of these terms we conclude
\begin{equation} \label{56term}
\lim_{x\rightarrow x'} \langle F_{\mu\nu} ( x) F^{\mu'\nu'} (x') \sigma_q \rangle =0.
\end{equation}
What remains therefore are the 1st and the 4th terms. Combining (\ref{1term}) and (\ref{2term}) and 
using (\ref{rel3} ) we conclude that 
\begin{equation}\label{ijcomp}
\langle T_{ij}(x) \rangle =\delta_{ij} \lim_{x\rightarrow x'} \widehat\nabla^2 G(x, x') .
\end{equation}

Let us now examine the transverse components of the stress tensor, we chose the free indices to be 
$a, b$. 
From (\ref{fullmaxstress}), we see that the first term is given by 
       \begin{align}\label{1tran}
           \nabla_{a}\nabla^{b'}G_{\lambda}^{~\lambda'}(x, x') &
           =\nabla_{a}\nabla^{b'}(G_{k}^{k'}(x, x') +G_{c}^{c'}(x, x') ),\nonumber\\
                                              &=\nabla_{a}\nabla^{b'}\left(-1-\frac{P_c P^{c'}}{\widehat\nabla^2}\right){G}(x, x') ,\nonumber\\
                                              &=\left( -\nabla_a\nabla^{b'}-\frac{\nabla_a P_c\nabla^{b'}P^{c'}}{\widehat\nabla^2}\right) {G}(x, x') .
       \end{align}
Expanding the second term we get
       \begin{align}\label{2tran}
      -\nabla_{a}^{~\lambda'}G_{\lambda }^{~b'}(x, x') =\frac{\nabla_a P_c \nabla^{c'}P^{b'}}{\widehat\nabla^2}     {G}(x, x') .
       \end{align}
 Similarly the third term of (\ref{fullmaxstress}) for the transverse components becomes
         \begin{align}\label{3tran}
     -\nabla^{b'}\nabla^{\lambda}G_{a\lambda'}(x, x') =\frac{\nabla_c P_a \nabla^{b'}P^{c'}}{\widehat\nabla^2}{G}(x, x') .
       \end{align}
 The fourth term results in 
       \begin{align}\label{4tran}
        \nabla_{\lambda}^{\lambda'}G_{a}^{~b'}(x, x') &=\left( -\nabla^2+\nabla_c\nabla^{c'}(\frac{-P_a P^{b'}}{\widehat\nabla^2}) \right) G(x, x') ,\nonumber\\
        &=\left( P_a P^{b'}-\frac{\nabla_c P_a \nabla^{c'} P^{b'}}{\widehat\nabla^2}\right) {G}(x, x') .
                                         \end{align}
The fifth and the sixth term together vanishes due to (\ref{56term}). Combining (\ref{1tran}), (\ref{2tran}), (\ref{3tran}) and (\ref{4tran}) we obtain
                                         \begin{align}
\langle T_a^{b}(x) \sigma_q\rangle &=\lim_{x\rightarrow x'} 
\left[-\nabla_a\nabla^{b'}+P_a P^{b'}+\frac{(\nabla_a P_c-\nabla_c P_a)(\nabla^{c'}P^{b'}-\nabla^{b'}P^{c'})}{\widehat\nabla^2}\right]{G}(x, x'),\nonumber\\
       &=\lim_{x\rightarrow x'} (-\nabla_a\nabla^{b'}+P_a P^{b'}- \delta_a^b \widehat\nabla^2) {G}( x, x'). 
                                         \end{align}
                                         To arrive at the last line we have used the identities in  (\ref{rel1}) and (\ref{rel2}).   We write down the non-trivial components explicitly
                                         \begin{eqnarray}
                                             \langle T_r^{r} (x) \sigma_q \rangle &=&
                                             \lim_{x\rightarrow x'} \left( -\partial_r\partial_{r'}+\frac{1}{r^2}\partial_{\theta}\partial_{\theta'}-\widehat\nabla^2\right) {G}(x, x') ,
                                              \\ \nonumber
    &=&  \frac{(q^2-1)  \left(11 q^2+1\right)}{720 \pi ^2 q^4 r^4}, \\ \nonumber
                                               \langle T_{\theta}^{\theta}(x)  \sigma_q \rangle
                                               &=& \lim_{x\rightarrow x'}  \left( \partial_r\partial_{r'}-\frac{1}{r^2}\partial_{\theta}\partial_{\theta'}-\widehat\nabla^2\right) {G} (x, x'),       \\ \nonumber
                                               &=& \frac{-11 q^4+10 q^2+1}{240 \pi ^2 q^4 r^4} .               
 \end{eqnarray}
 Transforming  these components to cartesian coordinates we get
 \begin{eqnarray}
 \langle T_{11}(x) \sigma_n\rangle&=&\frac{(q^2-1)  \left(11 q^2+1\right)}{720 \pi^2  q^4}\frac{(x_1^{2}-3x_2^{2})}{r^6},  \\ \nonumber
  \langle T_{22}(x)  \sigma_n\rangle&=&\frac{(q^2-1) \left(11 q^2+1\right)}{720 \pi^2  q^4}\frac{(x_2^{2}-3x_1^{2})}{r^6} , \\ \nonumber
  \langle T_{ij} (x) \sigma_n \rangle&=&-\delta_{ij}\frac{(q^2-1)  \left(11 q^2+1\right)}{720 \pi ^2 q^4 r^4}.
  \end{eqnarray}
  The last line results from taking the limit in (\ref{ijcomp}). 
  Now comparing the equations for the expectation value of the stress tensor in presence of the 
  co-dimension 2 defect in (\ref{stressdefhq}) we can read out the value of its conformal dimension
  \begin{equation}
  h_q = \frac{(q^2-1)  \left(11 q^2+1\right)}{360 \pi  q^3}.
  \end{equation}

\section{R\'{e}nyi entropies and  weights of twists} \label{appen3}

\subsection*{Symmetric tensors}
We list the R\'{e}nyi entropies and the conformal dimension of the co-dimensions two twist 
operator corresponding to the spherical entangling surface for all  $4\leq D\leq 14$ for 
symmetric rank-$s$ tensors.  

$\boldsymbol{D=4}$

\begin{eqnarray}
 S_q &=& \frac{ ( q+ 1) [ ( 30 s^2 + 1)q^2 + 1] }{180 q^3 } \log \epsilon.
 \end{eqnarray}
  \begin{eqnarray}
 h_q 
&=& \frac{( q^2 -1) ( q^2 ( 10 s^2 +1)  +1) }{ 360 \pi q^3}.
 \end{eqnarray}

$\boldsymbol{D=6}$

\begin{align}
   S_{q}=-\frac{(q+1) (s+1)^2 \left(\left(q^2+1\right) q^2 (7 s (s+2)+9)+2\right)}{30240 q^5 s!}\log(\epsilon).
\end{align}
\begin{align}
    h_{q}=\frac{(s+1)^2 \left(q^6 (21 s (s+2)+31)-21 q^2 (s+1)^2-10\right)}{151200 \pi ^2 q^5 s!}.
\end{align}

$\boldsymbol{D=8}$

\begin{align}
S_{q}&=\frac{(s+2) (s+3)!}{21772800 q^7 (1-q) s!}\Big[q^8 (-(18 s (s+4)+79))\nonumber\\
       &+14 q^4 (s+2)^2+4 q^2 (s (s+4)+5)+3\Big]\log(\epsilon).
\end{align}
\begin{align}
 h_{q}&= \frac{\left(q^2-1\right) (s+1) (s+2)^2 (s+3)}{50803200 \pi ^3 q^7} \Big[q^6 (62 s (s+4)+289)+q^4 (62 s (s+4)+289)\nonumber\\
         &+q^2 (20 s (s+4)+121)+21\Big].
 \end{align}
 
 $\boldsymbol{D=10}$
 
\begin{align}
 S_{q}&=\frac{-(s+3) (s+5)!}{1207084032000 q^9 (1-q) s!} \Big[q^{10} (-(2607 s (s+6)+24626))\nonumber\\
                                          &+1848 q^6 (s+3)^2+132 q^4 (5 s (s+6)+49)\nonumber\\
                                          &+99 q^2 (s (s+6)+14)+140\Big]\log(\epsilon).
\end{align}
\begin{align}
h_{q}&=\frac{\left(q^2-1\right) (s+3) \Gamma (s+6)}{301771008000 \pi ^4 q^9 s!}\Big[q^8 (3179 s (s+6)+31066)+q^6 (3179 s (s+6)\nonumber\\
         &+31066)+q^4 (1331 s (s+6)+14434)+21 q^2 (11 s (s+6)+174)+420\Big].
\end{align}

$\boldsymbol{D=12}$

\begin{align}
    S_{q}&=\frac{(s+4) (s+7)!}{79088145776640000 q^{11} (1-q) s!} \Big[q^{12} (-(320138 s (s+8)+5276673))\nonumber\\
            &+216216 q^8 (s+4)^2+1716 q^6 (49 s (s+8)+820)\nonumber\\
            &+9009 q^4 (2 s (s+8)+39)+1820 q^2 (s (s+8)+30)+4146\Big]\log(\epsilon).
\end{align}
\begin{align}
    h_{q}&=\frac{ \Gamma (s+8)(q^2-1) (s+4)}{4833164464128000 \pi ^5 q^{12} s!} \Big[q^2 (q^8 (403858 s (s+8)+6803477)\nonumber\\
            &+q^6 (403858 s (s+8)+6803477)+q^4 (187642 s (s+8)+3344021)\nonumber\\
            & +q^2 (47502 s (s+8)+998821)+5460 s (s+8)+179002)+15202\Big].
\end{align}

$\boldsymbol{D=14}$

\begin{align}
    S_{q}&=\frac{-(s+5) (s+9)! }{313189057275494400000 q^{13} (1-q) s!}\Big[q^{14} (-(1758891 s (s+10)+44865227))\nonumber\\
           & +1153152 q^{10} (s+5)^2+2288 q^8 (205 s (s+10)+5269)\nonumber\\
           &+429 q^6 (273 s (s+10)+7645)+1820 q^4 (10 s (s+10)+341)\nonumber\\
           & +1382 q^2 (s (s+10)+55)+4620\Big]\log(\epsilon).
\end{align}
\begin{align}
     h_{q}&=\frac{\left(q^2-1\right) (s+5) (s+9)!}{11309604846059520000 \pi ^6 q^{13} s!}\Big[q^{12} (6803477 s (s+10)+176203445)\nonumber\\
            &+q^{10} (6803477 s (s+10)+176203445)+q^8 (3344021 s (s+10)+89717045)\nonumber\\
            &+q^6 (998821 s (s+10)+29439685)+2 q^4 (89501 s (s+10)\nonumber\\
            &+3240875)+22 q^2 (691 s (s+10)+40735)+60060\Big].
\end{align}

We have verified that the above results satisfy the relation between  the derivatives of 
$S_q$ and $h_q$ found in \cite{Hung:2014npa}. The relation 
between the derivatives on $S_q$ and $h_q$ at $q= 1$ is given by
\begin{align} \label{myersrel}
    \partial_q S_q|_{q=1}=-\frac{ d{ \rm Vol} (AdS_{d}) }{2}\partial_q h_q|_{q=1}.
\end{align}
 For example in $D= 6$ we  see that 
 \begin{align}
      \partial_q S_q|_{q=1}&= -\frac{ 5{\rm Vol}(AdS_{d} )}{2}\partial_q h_q|_{q=1},\nonumber\\
                           &=\frac{(s+1)^2 (7 s (s+2)+12)}{5040 s!} \log\epsilon.
 \end{align}
 The relation (\ref{myersrel})  has been verified for all the listed values of $S_q$ and $h_q$.

\subsection*{$p$-forms}

We list the R\'{e}nyi entropies and the conformal dimensions of the co-dimension 2 twist operators
for conformal forms in even $D$ dimensions.  The conformal form is the $D/2-1$-form.

$\boldsymbol{D=4}:$ 1-form

\begin{eqnarray}
S_q 
 &=& \frac{(q+1)( 31 q^2  +1) }{ 180 q^3}  \log(\epsilon).
 \end{eqnarray}
\begin{eqnarray}
h_q 
&=& \frac{(q^2-1)( 11q^2+1) }{ 360 \pi q^3} .
\end{eqnarray}

$\boldsymbol{D=6}:$  2-form

 \begin{align}
      S_q=-\frac{(q+1) \left(877 q^4+37 q^2+2\right)}{5040 q^5}\log(\epsilon).
  \end{align}
   \begin{align}
      h_q=\frac{(q-1) (q+1) \left(191 q^4+23 q^2+2\right)}{5040 \pi ^2 q^5}.
  \end{align}

$\boldsymbol{D=8}: $ 3-form

\begin{align}
      S_q=\frac{(q+1) \left(15865 q^6+745 q^4+59 q^2+3\right)}{90720 q^7}\log(\epsilon).
  \end{align}
   \begin{align}
      h_q=\frac{(q-1) (q+1) \left(11 q^2+1\right) \left(227 q^4+10 q^2+3\right)}{30240 \pi ^3 q^7}.
  \end{align}

$\boldsymbol{D=10}:$  4-form

  \begin{align}
      S_q=-\frac{(q+1) \left(4200673 q^8+208993 q^6+19573 q^4+1555 q^2+70\right)}{23950080 q^9}\log(\epsilon).
  \end{align}
   \begin{align}
      h_q=\frac{(q-1) (q+1) \left(14797 q^8+2125 q^6+321 q^4+35 q^2+2\right)}{57024 \pi ^4 q^9}.
  \end{align}

$\boldsymbol{D=12}:$  5-form
 \begin{align}
      S_q &=\frac{(q+1)}{15567552000 q^{11}} \Big[2735714123 q^{10}+141122123 q^8\nonumber\\
               &+14539667 q^6+1420847 q^4+104246 q^2+4146\Big]\log(\epsilon).
  \end{align}
  \begin{align}
      h_q&=\frac{(q^2-1)}{86486400 \pi ^5 q^{11}} \Big[(92427157 q^{10}+13803157 q^8\nonumber\\
                &+2295661 q^6+307961 q^4+28682 q^2+1382\Big].
  \end{align}

  $\boldsymbol{D=14}: $ 6-form
  
   \begin{align}
      S_q &=-\frac{ (q+1)}{186810624000 q^{13}} \Big[32873434703 q^{12}+1738330703 q^{10}\nonumber\\
                &+190512431 q^8+21031119 q^6+1952202 q^4+130382 q^2+4620\Big]\log(\epsilon).
  \end{align}
   \begin{align}
      h_q &=\frac{(q-1) (q+1)}{6739200 \pi ^6 q^{13}} \Big[36740617 q^{12}+5636617 q^{10}\nonumber\\
                &+997801 q^8+151241 q^6+17822 q^4+1442 q^2+60\Big].
  \end{align}

\bibliographystyle{JHEP}
\bibliography{references} 
\end{document}